\documentclass[journal]{IEEEtran}
\pdfoutput=1
\usepackage[noadjust]{cite}
\usepackage[cmex10]{amsmath}
\usepackage{algorithmic}
\usepackage{array}
\usepackage{mdwmath}
\usepackage{eqparbox}
\usepackage{stfloats}
\usepackage{url}
\usepackage[pdftex]{graphicx}
\graphicspath{{../pdf/}{../jpeg/}}
\DeclareGraphicsExtensions{.pdf,.jpeg,.png}
\usepackage{epstopdf}
\usepackage{amsfonts}
\usepackage[tight,footnotesize]{subfigure}
\begin{document}

\title{3D Massive MIMO Systems: Modeling and Performance Analysis}

\author{Qurrat-Ul-Ain~Nadeem,~\IEEEmembership{Student Member,~IEEE,} Abla~Kammoun,~\IEEEmembership{Member,~IEEE,}
        M{\'e}rouane~Debbah,~\IEEEmembership{Fellow,~IEEE,} and ~ Mohamed-Slim~Alouini,~\IEEEmembership{Fellow,~IEEE}
\thanks{Manuscript received January 04, 2015; revised May 10, 2015; accepted July 16, 2015. }
\thanks{The work of Q.-U.-A. Nadeem, A. Kammoun and M. -S. Alouini was supported by a CRG 3 grant from the  Office of Sponsored Research at KAUST.  The work of M{\'e}rouane~Debbah was supported by ERC Starting Grant 305123 MORE (Advanced Mathematical Tools for Complex Network Engineering).}   
\thanks{Q.-U.-A. Nadeem, A. Kammoun and M.-S. Alouini are with the Computer, Electrical and Mathematical Sciences and Engineering (CEMSE) Division, King Abdullah University of Science and Technology (KAUST), Thuwal, Makkah Province, Saudi Arabia 23955-6900 (e-mail: \{qurratulain.nadeem,abla.kammoun,slim.alouini\}@kaust.edu.sa)}
\thanks{M. Debbah is  with  Sup{\'e}lec, Gif-sur-Yvette, France and Mathematical and Algorithmic Sciences Lab, Huawei France R\&D, Paris, France (e-mail: merouane.debbah@huawei.com, merouane.debbah@supelec.fr).}
}
\markboth{}%
{Shell \MakeLowercase{\textit{et al.}}: Bare Demo of IEEEtran.cls for Journals}
\maketitle
\begin{abstract}
Multiple-input-multiple-output (MIMO) systems of current LTE releases are capable of adaptation in the azimuth only. Recently, the trend is to enhance system performance by exploiting the channel's degrees of freedom in the elevation, which necessitates the characterization of 3D channels. We present an information-theoretic channel model for MIMO systems that supports the elevation dimension.  The model is based on the principle of maximum entropy, which enables us to determine the distribution of the channel matrix consistent with the prior information on the angles. Based on this model, we provide analytical expression for the cumulative density function (CDF) of the mutual information (MI) for systems with a single receive and finite number of transmit antennas in the general signal-to-interference-plus-noise-ratio (SINR) regime. The result is extended to systems with finite receive antennas in the low SINR regime. A Gaussian approximation to the asymptotic behavior of MI distribution is derived for the large number of transmit antennas and paths regime. We corroborate our analysis with simulations that study the performance gains realizable through meticulous selection of the transmit antenna downtilt angles, confirming the potential of elevation beamforming to enhance system performance. The results are directly applicable to the analysis of 5G 3D-Massive MIMO-systems.\end{abstract}

\section{Introduction}

Multiple-input multiple-output (MIMO) systems have remained a subject of interest in wireless communications over the past decade due to the significant gains they offer in terms of the spectral efficiency by exploiting the multipath richness of the channel. Pioneer work in this area has shown that for channel matrices with independent and identically distributed (i.i.d) elements, MIMO system capacity can potentially scale linearly with the minimum number of transmit (Tx) and receive (Rx) antennas \cite{Telatar}, \cite{foschini}. This has led to the emergence of massive MIMO systems, that scale up the conventional MIMO systems by possibly orders of magnitude compared to the current state-of-art \cite{massive}. These systems were initially designed to support antenna configurations at the base station (BS) capable of adaptation in the azimuth only. However, the future generation of mobile communication standards such as 3GPP LTE-Advanced are targeted to support even higher data rate transmissions, which has stirred a growing interest among researchers to enhance system performance through the use of adaptive electronic beam control over both the elevation and the azimuth dimensions. Several measurement campaigns have already demonstrated the significant impact of elevation on the system performance \cite{elevationcamp,elevationcamp1,practicals,ourwork}. The reason can be attributed to its potential to allow for a variety of strategies like user specific elevation beamforming and three-dimensional (3D) cell-splitting. This novel approach towards enhancing system performance uncovers entirely new problems that need urgent attention: deriving accurate 3D channel models, designing beamforming strategies that exploit the extra degrees of freedom, performance prediction and analysis to account for the impact of the channel component in the elevation and finding appropriate deployment scenarios. In fact, the 3GPP has started working on defining future mobile communication standards to help evaluate the potential of 3D beamforming \cite{TR36.873}. 

A good starting point for the effective performance analysis of  MIMO systems is the evaluation and characterization of their information-theoretic mutual information (MI). The statistical distribution of MI is very useful in obtaining performance measures that are decisive on the use of several emerging technologies. This has motivated many studies on the statistical distribution of the MI of 2D MIMO channels in the past decade \cite{wishart, wishart4, wishart3, wishart2,wishart1, gaussian2,gaussian,gaussian1}. Most of the work in this area assumes the channel gain matrix to have i.i.d complex Gaussian entries and resorts to the results developed for Wishart matrices. The complexity of the distribution of a Wishart matrix makes the analysis quite involved. The characteristic function (CF)/moment generating function (MGF) of the MI has been derived in \cite{gaussian1,correlatedcapacity,gaussian3} but the MGF approaches to obtain the MI distribution involve the inverse Laplace transform which leads to numerical integration methods. However in the large number of antennas regime, Random Matrix Theory (RMT) provides some simple deterministic approximations to the MI distribution. These deterministic equivalents were shown to be quite accurate even for a moderate number of antennas. The authors in \cite{gaussian2} showed that the MI can be well approximated by a Gaussian distribution and derived its mean and variance for spatially correlated Rician MIMO channels. More recently, Hachem \textit{et al} derived a Central Limit Theorem (CLT) for the MI of Kronecker channel model and rigorously proved that the MI converges to a standard Gaussian random variable in the asymptotic limit in \cite{walid_mutual_information}. More asymptotic results on the statistical distribution of MI of 2D channels have been provided in \cite{gaussian2,gaussian,gaussian1,gaussian3}. However, the extension of these results to the 3D case has not appeared yet. 

The efficient design and better understanding of the limits of wireless systems require accurate modeling and characterization of the 3D channels. Given the available knowledge on certain channel parameters like angle of departure (AoD), angle of arrival (AoA), delay, amplitude,  number of Tx and Rx antennas, finding the best way to model the channel consistent with the information available and to attribute a joint probability distribution to the entries of the channel matrix is of vital importance. It has been shown in literature that the choice of distribution with the greatest entropy creates a model out of the available information without making any arbitrary assumption on the information that was not available \cite{entropybook}, \cite{entropybook1}, \cite{entropy}. In the context of wireless communications, the authors in \cite{maruan} addressed the question of MIMO channel modeling from an information-theoretic point of view and provided guidelines to determine the distribution of the 2D MIMO channel matrix through an extensive use of the principle of maximum entropy together with the consistency argument. Several channel models were derived consistent with the apriori knowledge on parameters like AoA, AoD, number of scatterers. These results can be extended to the 3D MIMO channel model, which will be one of the aims of this work. As mentioned earlier, almost all existing channel models are 2D. However, owing to the growing interest in 3D beamforming, extensions of these channel models to the 3D case have started to emerge recently \cite{Winner+}, \cite{TR36.873}. An important feature of these models is the vertical downtilt angle, that can be optimized to change the vertical beam pattern dynamically and  yield the promised MI gains \cite{tilt,utility,antennatilt,caire}.  Studies on 3D channel modeling and impact of antenna downtilt angle on the achievable rates have started to appear in literature, but to the best of authors' knowledge, there has not been any comprehensive study to develop analytical expressions for the MI of these channels, that could help evaluate the performance of the future 3D MIMO systems. 

The aims of this paper are twofold. First is to provide an information-theoretic channel model for 3D massive MIMO systems and second is to predict and analyze the performance of these systems by characterizing the distribution of the MI. We follow the guidelines provided in \cite{maruan} to present the entropy maximizing 3D channel model that is consistent with the state of available knowledge of the channel parameters. The 3D channel used for the MI analysis is inspired from the models presented in standards like 3GPP SCM \cite{SCM}, WINNER+ \cite{Winner+} and ITU \cite{ITU}. These standardized MIMO channel models assume single-bounce scattering between the transmitter and receiver, which allows us to assume the AoDs and AoAs to be fixed and known apriori during the modeling phase. The maximum entropy 3D channel model consistent with the state of knowledge of channel statistics pertaining to AoDs and AoAs, turns out to have a systematic structure with a reduced degree of randomness in the channel matrix. This calls for the use of an approach that does not depend on the more generally employed results on the distribution of Wishart matrices. However, the systematic structure of the model can be exploited to our aid in characterizing its MI. With this channel model at hand, we derive analytical expressions for the cumulative density function (CDF) of the MI in different signal-to-interference plus noise ratio (SINR) regimes. An exact closed-form expression is obtained for systems with a single Rx and multiple Tx antennas in the general SINR regime. The result is extended to systems with a finite number of Rx antennas in the low SINR regime. We also present an asymptotic analysis of the statistical distribution of the MI and show that it is well approximated by a Gaussian distribution for any number of Rx antennas, as the number of paths and Tx antennas grow large. Finally numerical results are provided that illustrate an excellent fit between the Monte-Carlo simulated CDF and the derived closed-form expressions. The simulation results provide a flavor of the performance gains realizable through the meticulous selection of the downtilt angles.  We think our work contributes to cover new aspects that have not yet been investigated by a research publication and addresses some of the current challenges faced by researchers and industrials to fairly evaluate 3D beamforming techniques on the basis of achievable rates. The results have immediate applications in  the design of 3D 5G massive MIMO systems.

The rest of the paper is organized as follows. In section II, we introduce the 3D channel model based on the 3GPP standards and information-theoretic considerations. In Section III, we provide analytical expressions for the CDF of the MI in the general and low SINR regimes for a finite number of Tx antennas. In section IV, we provide an asymptotic analysis that holds for any number of Rx antennas. The derived results are corroborated using simulations in section V and finally in section VI, some concluding remarks are drawn.

\section{System and Channel Model}

Encouraged by the potential of elevation beamforming to enhance system performance, many standardized channel models have started to emerge that define the next generation 3D channels. We base the evaluation of our work on these models while making some realistic assumptions on the channel parameters. 

\subsection{System Model}
\begin{figure}[!b]
\centering
\includegraphics[width=2.25in]{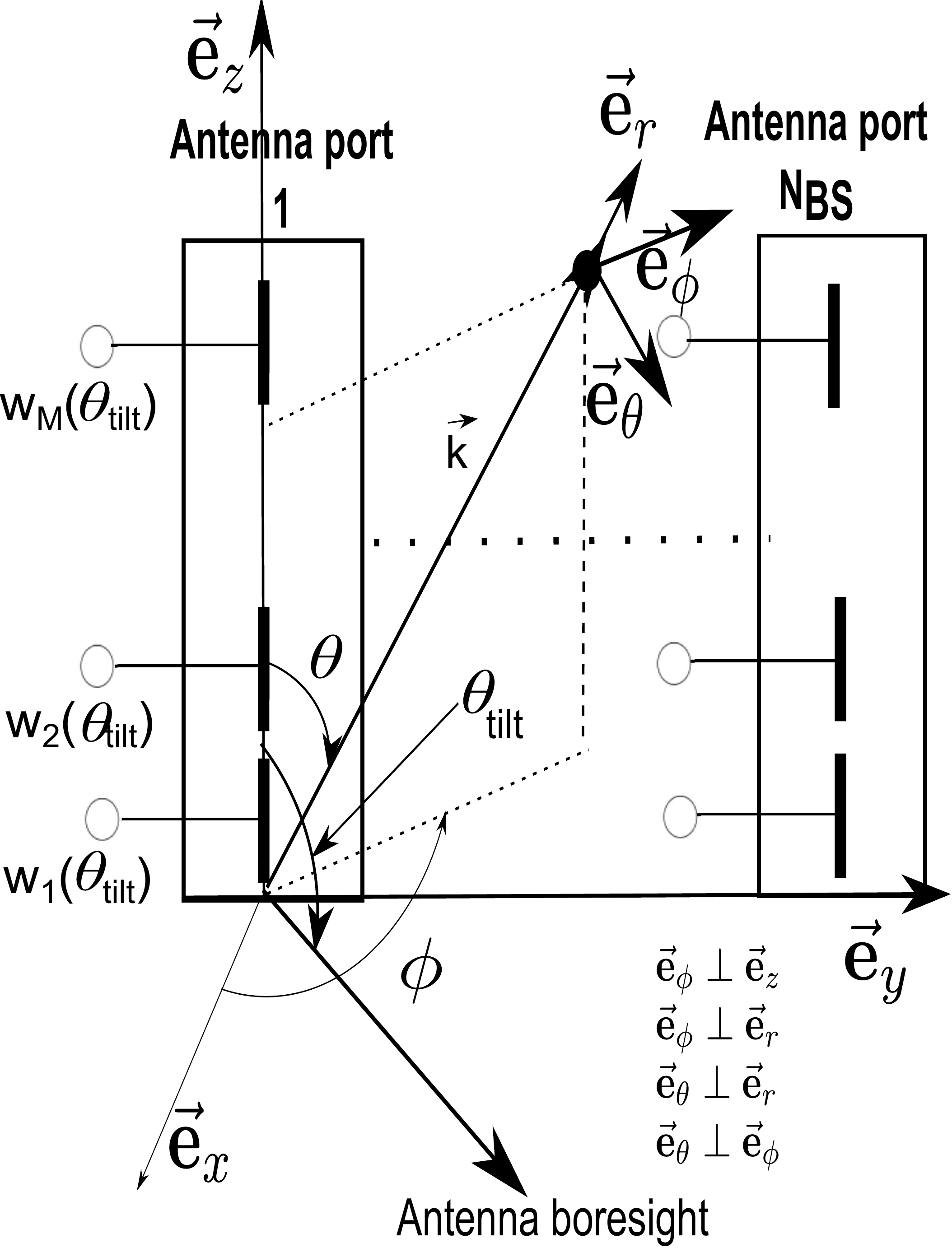}
\caption{Antenna configuration.}
\label{antennacon}
\end{figure}

We consider a downlink 3D MIMO system, where the BS is equipped with $N_{BS}$ directional Tx antenna ports and the mobile station (MS) has $N_{MS}$ Rx antenna ports. The antenna configuration is shown in Fig.\ref{antennacon}. Note that $\theta_{tilt}$ represents the elevation angle of the antenna boresight, a parameter that does not appear in the 2D channel models. Also, zero electrical downtilt corresponds to $\theta_{tilt}=\frac{\pi}{2}$.  In LTE, the radio resource is organized on the basis of antenna ports. Each antenna port is mapped to a group of physical antenna elements which carry the same signal and constitute an antenna. The spatial multiplexing is performed across the ports.

\begin{figure*}[!t]
\begin{align}
\label{channel}
\text{[\textbf{H}]}_{su}&=\frac{1}{N}\sum\limits_{n=1}^N  \alpha_{n} \sqrt{g_{t}(\phi_{n},\theta_{n},\theta_{tilt})} \exp \left(i k (s-1) d_{t} \sin\phi_{n}\sin\theta_{n}\right) \sqrt{g_{r}(\varphi_{n},\vartheta_{n})}  \exp\left(i k  (u-1) d_{r}\sin\varphi_{n}\sin\vartheta_{n}\right),
\end{align}
\hrulefill
\vspace*{4pt}
\end{figure*}

In the configuration shown in Fig. \ref{antennacon}, there are $N_{BS}$ antenna ports placed in the horizontal plane, along the $\hat{\textbf{e}}_{y}$ direction. Each antenna port is mapped to $M$ vertically stacked antenna elements that determine the effective port pattern. The same Tx signal is fed to all elements in a port with corresponding weights $w_{k}$'s, $k=1, \dots, M$, in order to achieve the desired directivity. In the current activity around 3D channel modeling by the 3GPP, the proposed method for 3D beamforming is to adjust the weights applied to the elements in a port to obtain the desired downtilt angle for that port [8]. The same is done for the other ports. The MS sees each antenna port as a single antenna because all the elements carry the same signal. Therefore, we are interested in the channel between the transmitting antenna port and the receiver side.

\subsection{3D MIMO Channel Model}

The mobile communication standards like 3GPP SCM \cite{SCM}, ITU \cite{ITU} and WINNER \cite{Winner}  follow a system level, stochastic channel modeling approach wherein, the propagation paths are described through statistical parameters, like delay, amplitude, AoA and AoD. These models generate channel realizations by summing the contributions of $N$ multipaths (plane waves) with specific parameters.  In the existing 2D channel models like the 3GPP SCM and WINNER, the propagation paths are described using azimuth angles alone. Moreover, the value of the elevation angle of the antenna boresight ($\theta_{tilt}$) is assumed to be fixed at $\frac{\pi}{2}$ in these 2D models. However, as shown by several measurement campaigns \cite{elevationcamp,elevationcamp1}, there is a significant component of energy that is radiated in the elevation so characterizing the propagation paths in the azimuth alone is not a true depiction of the environment. Also, assuming the elevation angle of the antenna boresight to be fixed implies that the channel's degrees of freedom in the elevation are not being exploited. The dynamic adaptation of the downtilt angles can open up several possibilities for 3D beamforming that can bring about significant performance gains. Therefore the extension of the existing 2D models to the 3D case needs to take into account the elevation angles of the propagation paths and introduce the parameter $\theta_{tilt}$ into the expressions of the antenna pattern rather than assuming it to be fixed. These extensions have started to emerge recently \cite{Winner+}, \cite{3GPP3D}, \cite{ITU}.  

\begin{figure}[!b]
\centering
\includegraphics[width=3.25in]{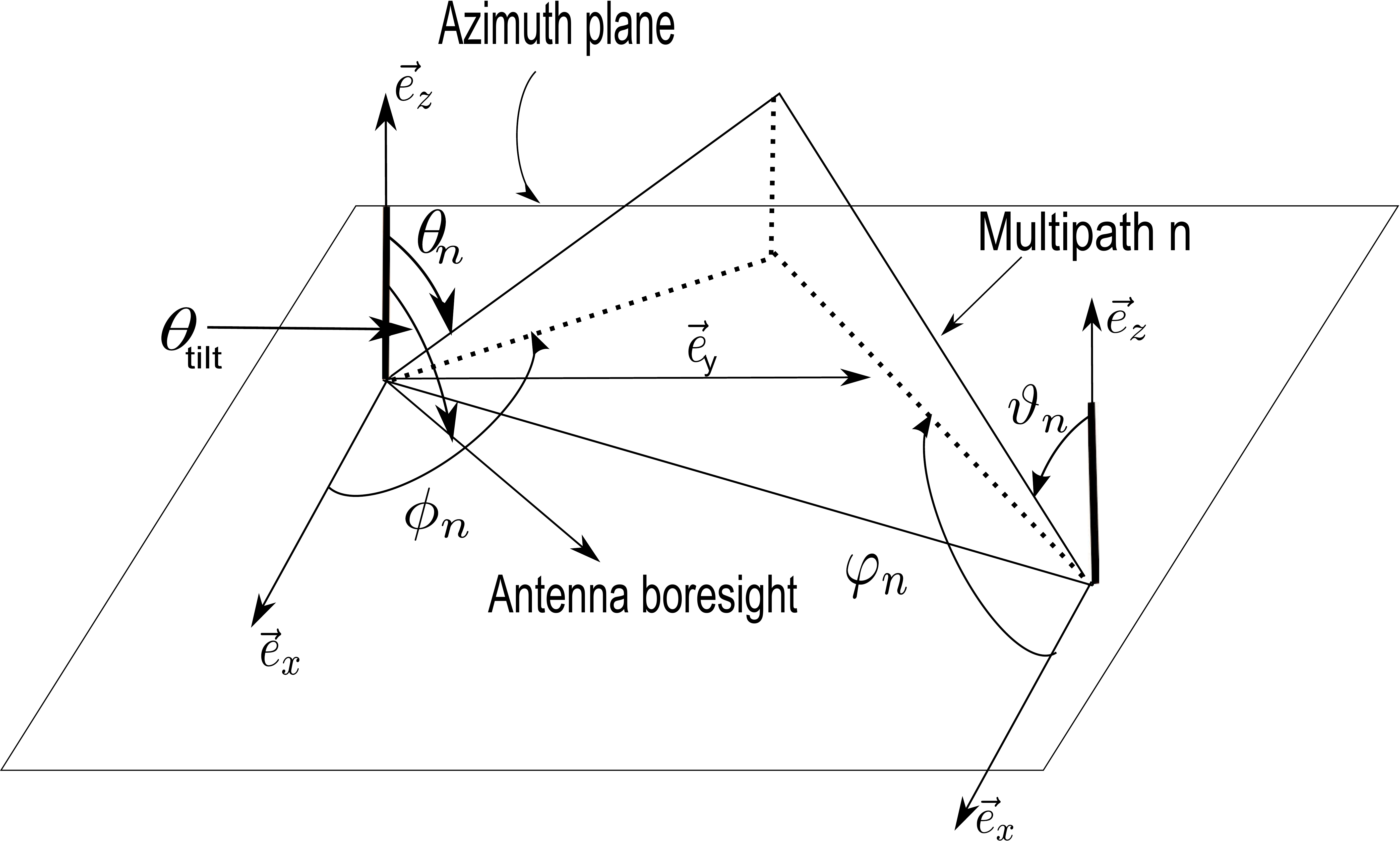}
\caption{3D channel model.}
\label{3dchannel}
\end{figure}

Based on the aforementioned standards and for the antenna configuration in Fig. \ref{antennacon}, the effective 3D channel between the BS antenna port $s$ and the MS antenna port $u$ is given by (\ref{channel}), where  $\phi_{n}$ and $\theta_{n}$ are the azimuth and elevation AoD of the $n^{th}$ path respectively, $\varphi_{n}$ and $\vartheta_{n}$ are the azimuth and elevation AoA of the $n^{th}$ path respectively and $\theta_{tilt}$ is the elevation angle of the antenna boresight. ${\alpha}_{n}$ is the complex random amplitude of the $n^{th}$ path. Also $\sqrt{\text{g}_{t}(\phi_{n},\theta_{n},\theta_{tilt})}$ and $\sqrt{\text{g}_{r}(\varphi_{n},\vartheta_{n})}$ are the global patterns of Tx and Rx antennas respectively. $d_{t}$ and $d_{r}$ are the separations between Tx antenna ports and Rx antenna ports respectively and  $k$ is the wave number that equals $\frac{2\pi}{\lambda}$, where $\lambda$ is the wavelength of the carrier frequency. The entries  \cite{drabla},
\begin{eqnarray}
{[\textbf{a}_{t}(\phi, \theta)]}_{s}&{}={}&\exp \left(ik (s-1) d_{t} \sin\phi\sin\theta\right), \\
{[\textbf{a}_{r}(\varphi, \vartheta)]}_{u}&{}={}&\exp\left(ik (u-1) d_{r}\sin\varphi \sin\vartheta\right)
\end{eqnarray}
are the array responses of $s^{th}$ Tx and $u^{th}$ Rx antenna respectively. Fig. \ref{3dchannel} shows the 3D channel model being considered. 

To enable an abstraction of the role played by antenna elements in performing the downtilt, ITU approximates the global pattern of each port by a narrow beam as \cite{ITU}, \cite{drabla},
\begin{align}
&\sqrt{[17\rm{dBi}-\text{min}\{-(A_{H}(\phi)+A_{V}(\theta,\theta_{tilt})),20 \rm{dB}\}]_{\text{lin}}},
\end{align}
where,
\begin{align}
A_{H}(\phi)&= -\text{min}\left[ 12 \left(\frac{\phi}{\phi_{3dB}}\right)^2, 20 \right] \rm{dB}, \\
A_{V}(\theta,\theta_{tilt})&= -\text{min} \left[12 \left(\frac{\theta-\theta_{tilt}}{\theta_{3dB}}\right)^2,20 \right] \rm{dB}.
\end{align}  
where $\phi_{3dB}$ is the horizontal 3 $\rm{dB}$ beamwidth and $\theta_{3dB}$ is the vertical 3 $\rm{dB}$ beamwidth. The individual antenna radiation pattern at the MS $g_{r}(\varphi_{n},\vartheta_{n})$ is taken to be $0$ $\rm{dB}$. Note that the expressions for the radiation pattern of the antenna ports in the 3D model include the parameter $\theta_{tilt}$ to allow for the dynamic adaptation of the elevation angle of the antenna boresight, as discussed earlier. 

\subsection{Maximum Entropy Channel}

The model just presented based on the industrial standards is a propagation-motivated model that explicitly sums up the contribution of several multipaths. The theoretical analysis of this 3D channel model is difficult due to several reasons. The multiple propagation paths result in a large number of random variables (RVs) in the model. Secondly, the model exhibits non-linearity with respect to the RVs, AoDs and AoAs, that appear as arguments of the exponential terms. In order to facilitate the theoretical analysis, this model can be reformulated to develop an equivalent analytical channel model that is also propagation-motivated but has a more compact structure. In this work, we use the principle of maximum entropy to bridge the gap between standards and theory and develop an equivalent information-theoretic analytical model for (\ref{channel}), which will circumvent the problems just mentioned and facilitate the development of closed-form expressions for the MI in the subsequent sections.

 It was rigorously proved in \cite{entropy} that the principle of maximum entropy yields models that express the constraints of our knowledge of model parameters and avoid making any arbitrary assumptions on the information that is not available. Maximizing any quantity other than entropy would result in inconsistencies, and the derived model will not truly reflect our state of knowledge. Motivated by the success of the principle of maximum entropy in parameter estimation and Bayesian spectrum analysis modeling problems, the authors in \cite{maruan} utilized this framework to devise theoretical grounds for constructing channel models for 2D MIMO systems, consistent with the state of available knowledge of channel parameters. 

The geometry-based MIMO channel models presented in standards assume single-bounce scattering between the transmitter and receiver, i.e., there is a one-to-one mapping between the AoDs and AoAs, allowing several bounces to occur as long as each AoA is linked to only one AoD \cite{maruansurvey}. Moreover the positions of scatterers, BS and MS are assumed to be fixed and known to the modeler of the channel. Consequently, the modeler can compute the AoDs and AoAs and $\phi_{n},\theta_{n},\varphi_{n},\vartheta_{n}$ are therefore assumed to be known apriori and fixed over channel realizations. Another motivation behind assuming the angles to be known apriori was provided in \cite{maruan}, where the authors use the observation that for single-bounce scattering channel models the directions of arrival and departure are deterministically related by Descartes's laws, to assume that the channel statistics pertaining to AoD and AoA do not change during the modeling phase. The apriori knowledge of these parameters helps us resolve the problem of the non-linearity in (\ref{channel}), by conditioning on the RVs that exhibit this non-linear dependence with the model. The only random component in the 3D channel model is now $\alpha_{n}$, so a suitable distribution needs to be assigned to it such that the obtained model is consistent with the information already available. 

It was proved through an extensive analysis in \cite{maruan} that in the presence of the prior information on the number of propagation paths, AoDs, AoAs and transmitted and received power of the propagation paths, the maximum entropy channel model is given by,
\begin{align}
\label{maruanchannel}
\textbf{H}=\boldsymbol{\Psi} \textbf{P}_{Rx}^{\frac{1}{2}} (\boldsymbol{\Omega} \circ \textbf{G}) \textbf{P}_{Tx}^{\frac{1}{2}} \boldsymbol{\Phi}^{H},
\end{align}
where $\textbf{P}_{Rx}$ and $\textbf{P}_{Tx}$ contain the corresponding transmitted and received powers of the multipaths, $\boldsymbol{\Omega}$ is the mask matrix  that captures the path gains, $\boldsymbol{\Psi}$ and $\boldsymbol{\Phi}$ capture the antenna array responses and $\circ$ is the Hadamard product. The solution of the consistency argument that maximizes the entropy is to take the unknown random matrix \textbf{G} to be i.i.d zero mean Gaussian with unit variance \cite{maruan}, \cite{maruansurvey}. 

To represent the channel model in (\ref{channel}) with a structure similar to that of (\ref{maruanchannel}), we define \textbf{A} and \textbf{B} as $N_{BS}\times N$ and $N_{MS}\times N$  deterministic matrices given by,
\begin{align}
\label{A}
& \textbf{A}_{}=\frac{1}{\sqrt{N}}[\textbf{a}_{t}(\phi_{1},\theta_{1}), \textbf{a}_{t}(\phi_{2},\theta_{2}), ......, \textbf{a}_{t}(\phi_{N},\theta_{N})] \circ \\
& \begin{bmatrix}
\ \sqrt{g(\phi_{1},\theta_{1},\theta_{tilt_{1}})}   & \dots       & \sqrt{g(\phi_{N},\theta_{N},\theta_{tilt_{1}})}     \\
\vdots & \ddots & \vdots   \\
\ \sqrt{g(\phi_{1},\theta_{1},\theta_{tilt_{N_{BS}}})}    & \dots       & \sqrt{g(\phi_{N},\theta_{N},\theta_{tilt_{N_{BS}}})} \end{bmatrix}, \nonumber
\end{align}
where $\textbf{a}_{t}(\phi_{n},\theta_{n})$=$[[\textbf{a}_{t}(\phi_{n}, \theta_{n})]_{s=1},...,[\textbf{a}_{t}(\phi_{n}, \theta_{n})]_{s=N_{BS}}]^{H}$. Similarly,
\begin{align}
\label{B}
\textbf{B}&= [\textbf{a}_{r}(\varphi_{1},\vartheta_{1}), ...., \textbf{a}_{r}(\varphi_{N},\vartheta_{N})],
\end{align}
where $\textbf{a}_{r}(\varphi_{n},\vartheta_{n})=[[\textbf{a}_{r}(\varphi_{n}, \vartheta_{n})]_{u=1},...,[\textbf{a}_{r}(\varphi_{n}, \vartheta_{n})]_{u=N_{MS}}]^{T}$.

Note that the array responses and antenna patterns are captured in \textbf{A} and \textbf{B}. Since the transmitted and received powers are incorporated in the antenna patterns so $\textbf{P}_{Rx}$ and $\textbf{P}_{Tx}$ in (7) are identity matrices. For the single-bounce scattering model, the mask matrix $\boldsymbol{\Omega}$ will be diagonal. Taking $\boldsymbol{\Psi}$ and $\boldsymbol{\Phi}$ to be \textbf{B} and \textbf{A} respectively, the solution of the maximum entropy problem for the $N_{MS} \times N_{BS}$ single-bounce scattering 3D MIMO channel model in (\ref{channel}) will have a systematic structure as follows,
\begin{equation}
\label{entrop}
\begin{aligned}
\textbf{H}=\frac{1}{\sqrt{N}}\textbf{B} \hspace{.03 in} \text{diag}(\boldsymbol{\alpha}) \hspace{.02 in} \textbf{A}^{H},
\end{aligned}
\end{equation}
where $\boldsymbol{\alpha}$ is an $N$ dimensional vector with entries that are i.i.d zero mean, unit variance Gaussian RVs.

\section{Mutual Information Analysis in Finite Number of Antennas Regime}

In this section, we explicitly define the MI for subsequent performance prediction and analysis. We derive an analytical expression for the statistical distribution of the MI in the general SINR regime for a single Rx antenna and any arbitrary finite number of Tx antennas. The result is then extended to an arbitrary finite number of Rx antennas in the low SINR regime. 

We consider a point-to-point communication link where \textbf{H} is a $N_{MS} \times N_{BS}$ MIMO channel matrix from (\ref{entrop}). The channel is linear and time-invariant. The received complex baseband signal $\textbf{y} \in \mathbb{C}^{N_{MS} \times 1} $ at the MS is given by,
\begin{align}
\textbf{y}=\textbf{H} \textbf{s}+\textbf{n},
\end{align}
where $\textbf{s} \in \mathbb{C}^{N_{BS} \times 1}$ is the Tx signal from the BS and $\textbf{n} \in \mathbb{C}^{N_{MS} \times 1}$ is the complex additive noise such that $\mathbb{E}[\textbf{n} \textbf{n}^{H}] = \textbf{R}+\sigma^{2}\textbf{I}$, where $\textbf{R}$ is the covariance matrix of interference experienced by the MS and $\sigma^{2}$ is the noise variance at the MS.
 
In this context, the MI of the 3D MIMO system is given by,
\begin{equation}
\label{MI}
\begin{aligned}
I(\sigma^2)=\log \det (\textbf{I}_{N_{MS}}+ (\textbf{R}+\sigma^2 \textbf{I}_{N_{MS}})^{-1} \textbf{H} \textbf{H}^{H}).
\end{aligned}
\end{equation}

\textbf{H} is fixed during the communication interval, so we do not need to time average the MI.

\subsection{General SINR Regime}

The exact MI given in (\ref{MI}) has been mostly characterized in literature for the cases when $\textbf{H}\textbf{H}^{H}$ is a Wishart matrix  \cite{wishart,wishart1,wishart4,wishart3}. Wishart matrices are random matrices with independent entries and have been frequently studied in the context of wireless communications. However the special structure of our channel model reduces the degree of randomness as now \textbf{H} is a diagonal matrix of random Gaussian RVs as opposed to the whole matrix having random entries, which implies that we can not employ the results developed for Wishart matrices. However fortunately for the single-bounce scattering model in (\ref{entrop}), we can obtain results by exploiting the quadratic nature of the entries of $\textbf{H}\textbf{H}^{H}$. For the case, when MS is equipped with a single antenna, an exact closed-form solution can be obtained and is provided in the following theorem.

\textit{Theorem 1:} Let the MS be equipped with a single Rx antenna and the BS with any arbitrary number of Tx antennas, then the statistical distribution of MI in the general SINR regime can be expressed in a closed-form as,
\begin{align}
\label{Theorem2}
\mathbb{P}[I(\sigma^2) \leq y ]&=1-\sum\limits_{i=1}^N  \frac{\lambda_{i}^{N}}{\prod\limits_{l\neq i}^N (\lambda_{i}-\lambda_{l})}\frac{1}{\lambda_{i}} \exp \left(-\frac{(e^{y}-1)}{\lambda_{i}}\right),
\end{align}
where $\lambda_{i}$'s are the eigenvalues of \textbf{C} given that,
\begin{align}
\label{C_gen}
\textbf{C}=\frac{1}{N}(\textbf{B}^{H} \boldsymbol{\Omega} \textbf{B}) \circ (\textbf{A}^{H}  \textbf{A})^{T},
\end{align}
where $\boldsymbol{\Omega}=(\textbf{R}+ \sigma^{2}\textbf{I}_{MS})^{-1}$.

\textit{Proof:} The proof follows from realizing that when the number of antennas at the MS is one, $(\textbf{R}+\sigma^2 \textbf{I}_{N_{MS}})^{-1} \textbf{H}\textbf{H}^{H}$ is a scalar and can be expressed as a quadratic form in $\boldsymbol{\alpha}$. The MI is given as a function of this quadratic form as,
\begin{align}
I(\sigma^2)&=\text{log }(1+ {\boldsymbol{\alpha}}^{H} \textbf{C} \boldsymbol{\alpha}).
\end{align}
where \textbf{C} is given by (\ref{C_gen}). 

This quadratic form in Normal RVs will play a crucial role in the subsequent analysis in this work. The CF of a quadratic form in Normal RVs is given by \cite{quadraticchar1}, \cite{quadraticchar},
\begin{align}
\label{char}
\mathbb{E}[e^{j({\boldsymbol{\alpha}}^{H} \textbf{C} \boldsymbol{\alpha})}]&= \frac{1}{\det[\textbf{I}_{N}-j \frac{1}{N}((\textbf{B}^{H} \boldsymbol{\Omega} \textbf{B}) \circ (\textbf{A}^{H} \textbf{A})^{T})]}.
\end{align}

It is easy to see that the denominator of (\ref{char}) can be equivalently expressed as $\prod_{i=1}^{N} (1-j\lambda_{i})$, where $\lambda_{i}$'s are the eigenvalues of \textbf{C}. (\ref{char}) has a form similar to the CF of the sum of i.i.d unit parameter exponential RVs scaled by $\lambda_{i}$'s, i.e. ${\boldsymbol{\alpha}}^{H} \textbf{C} \boldsymbol{\alpha} \sim \sum_{i=1}^{N} \lambda_{i} \text{Exp}(1)$.  Consequently, the closed-form expression of the statistical distribution of $(\textbf{R}+\sigma^2 \textbf{I}_{N_{MS}})^{-1} \textbf{H} \textbf{H}^{H}$ is given by \cite{expdist},
\begin{equation}
\begin{aligned}
\mathbb{P}[{\boldsymbol{\alpha}}^{H} \textbf{C} \boldsymbol{\alpha} \leq x ]=1-\sum\limits_{i=1}^N  \frac{\lambda_{i}^{N}}{\prod\limits_{l\neq i}^N (\lambda_{i}-\lambda_{l})}\frac{1}{\lambda_{i}}\exp\left(-\frac{x}{\lambda_{i}}\right),  \nonumber
\end{aligned}
\end{equation}

The rest of the analysis can be completed through the transformation $\text{log}(1+x)$. 
\begin{align}
\mathbb{P}[I(\sigma^2) \leq y]=\mathbb{P}[\text{log}(1+x) \leq y ]= F_{x}(e^{y}-1).
\end{align}

The theoretical result coincides very well with the simulated CDF for the general SINR regime, as will be illustrated later in the simulations section. 

\subsection{Low SINR Regime}
It is well known that the performance of the channel can severely deteriorate in the low SINR regime, unless intelligent precoding and signaling schemes are employed \cite{peaky}. It is important to develop analytical expressions for the MI of MIMO channels in this regime to allow for the analysis of various signaling and precoding schemes. We develop such an analytical expression for the 3D channel model in (\ref{entrop}) by exploiting its structure. The closed-form expression for the CDF of the MI in the low SINR regime is now presented in the following theorem.

\textit{Theorem 2:} In the low SINR regime, the approximate statistical distribution of the MI can be expressed in a closed-form as,
\begin{equation}
\label{Theorem1}
\begin{aligned}
\mathbb{P}[I(\sigma^2) \leq x ]=1-\sum\limits_{i=1}^N  \frac{\lambda_{i}^{N}}{\prod\limits_{l\neq i}^N (\lambda_{i}-\lambda_{l})}\frac{1}{\lambda_{i}}\exp\left(-\frac{x}{\lambda_{i}}\right),
\end{aligned}
\end{equation}
where $\lambda_{i}$'s are the eigenvalues of \textbf{C} given in (\ref{C_gen}).

\textit{Proof:} The proof of Theorem 2 follows from making use of the relation that for $\boldsymbol{\delta}$ with $||\delta||_{2} \approx 0$,
\begin{align}
\label{approx}
\log \det(\textbf{X} + \boldsymbol{\delta}\textbf{X}) &= \log \det(\textbf{X}) + Tr(\textbf{X}^{-1}\boldsymbol{\delta} \textbf{X}).
\end{align}
In the low SINR regime, $||(\textbf{R}+ \sigma^{2}\textbf{I}_{N_{MS}})^{-1}||_{2} \approx 0$ so taking $\boldsymbol{\delta}$ as $(\textbf{R}+ \sigma^{2}\textbf{I}_{N_{MS}})^{-1} \textbf{H} \textbf{H}^{H}$ and \textbf{X} as $\textbf{I}_{N_{MS}}$ yields,
\begin{align}
\label{MIlow}
I(\sigma^2)&=Tr((\textbf{R}+ \sigma^{2}\textbf{I}_{N_{MS}})^{-1}\textbf{H}\textbf{H}^{H}).
\end{align}

Observe from (\ref{entrop}) that the trace of $\textbf{H}\textbf{H}^{H}$ can be expressed as a quadratic form given by,
\begin{align}
\label{quad}
Tr(\boldsymbol{\Omega} \textbf{H}\textbf{H}^{H})&=\mathbf{\boldsymbol{\alpha}}^{H} \textbf{C} \boldsymbol{\alpha},
\end{align}
where $\boldsymbol{\Omega}$ is an arbitrary deterministic matrix and 
\begin{align}
{[\textbf{C}]}_{ij}&=\frac{1}{N}[\textbf{B}^{H} \boldsymbol{\Omega} \textbf{B}]_{ij} \circ [\textbf{A}^{H}\textbf{A}]_{ji}.
\end{align}

Using (\ref{quad}), $I(\sigma^{2})$ in (\ref{MIlow}) is a quadratic form in zero mean, unit variance Normal i.i.d RVs $\alpha$'s,
\begin{align}
I(\sigma^{2})&=\boldsymbol{\alpha}^{H} \textbf{C} \boldsymbol{\alpha},
\end{align}
where \textbf{C} is given by (\ref{C_gen}). The rest of the proof follows from that for the general SINR regime, where we already characterized the distribution of 
${\boldsymbol{\alpha}}^{H} \textbf{C} \boldsymbol{\alpha}$. Numerical results presented in section V verify this expression. 

\section{Asymptotic Analysis of the Mutual Information}
As explained earlier, the reduced degree of randomness in our model makes it difficult to characterize the behaviour of the MI. The case for a single Rx antenna was dealt with in the last section and a closed-form was obtained as a consequence of the quadratic nature of the entries of $\textbf{H}\textbf{H}^{H}$. Even when $\textbf{H}\textbf{H}^{H}$ is a Wishart matrix, the study of the exact statistical distribution of the MI becomes rather involved because it often succumbs to taking the inverse Fourier transform of the CF or applying Jacobian transformations on the eigenvalue distribution of Wishart matrices \cite{wishart2}, \cite{wishart1}. As a result, a lot of work resorts to the asymptotic analysis and provides Gaussian approximations to the distribution of the MI. The asymptotic analysis provided here differs from the previous analysis because of the non-Wishart nature of $\textbf{H}\textbf{H}^{H}$ and the incorporation of the elevation dimension and antenna tilt angles in the channel model.

In order to study the asymptotic approximation to the MI distribution, the following assumption is required over the number of BS antennas and multipaths. 
\\
\textbf{Assumption A-1.} In the large $(N_{BS}, N)$ regime, $N_{BS}$ and $N$ tend to infinity such that
\begin{align}
\label{A1}
0 < \lim \inf \frac{N_{BS}}{N} \leq \lim \sup \frac{N_{BS}}{N} < + \infty,
\end{align}
a condition we shall refer to by writing $N_{BS}$, $N \rightarrow \infty$. This assumption basically implies that for the analysis to hold, $N$ and $N_{BS}$ have to be of comparable orders. This is valid in practice too as standards like SCM and ITU assume the total number of resolvable and unresolvable paths to be around $100$ depending on the scenario. This is also a reasonable value for the number of antennas in massive MIMO systems. In fact, this assumption means that to reap the benefits of deploying multiple antennas at the BS, the number of propagation paths should be of the same order as the number of antennas, otherwise it would result in what is known as the pinhole effect \cite{ourwork}. $N$ and $N_{BS}$ need to grow large to allow us to use tools from random matrix theory, but the analysis is valid for finite system dimensions as well as will be confirmed through the simulation results in section V.

\subsection{Distribution of \textbf{H}$\textbf{H}^{H}$}
The analysis starts by characterizing the distribution of \textbf{H}$\textbf{H}^{H}$, which would later be followed by transformations to complete  the characterization of the MI. Defining $\bar{\textbf{b}}_{i}$ as the $i^{th}$ row vector of \textbf{B} in (\ref{B}), the $(k,l)^{th}$ entry of \textbf{H}$\textbf{H}^{H}$ can be expressed as,
\begin{equation}
\begin{aligned}
{[\textbf{H}\textbf{H}^{H}]}_{kl}=\frac{1}{N} \boldsymbol{\alpha}^{H} ([\bar{\textbf{b}}_{k}^{H} \bar{\textbf{b}}_{l}]\circ (\textbf{A}^{H} \textbf{A})^{T}) \boldsymbol{\alpha}.
\end{aligned}
\end{equation}

To characterize the distribution, an additional assumption is required over the matrices that form the quadratic forms in \textbf{H}$\textbf{H}^{H}$. Denoting $N_{MS}$ by $M$, the assumption is stated as follows. 
\\
\textbf{Assumption A-2.} 
Under the setting of Assumption \textbf{A-1}, 
\begin{align}
\label{A2}
 \frac{||[\bar{\textbf{b}}_{k}^{H} \bar{\textbf{b}}_{l}]\circ (\textbf{A}^{H} \textbf{A})^{T}||_{sp}}{||[\bar{\textbf{b}}_{k}^{H} \bar{\textbf{b}}_{l}]\circ (\textbf{A}^{H} \textbf{A})^{T}||_{F} } \xrightarrow{} 0, \hspace{.08 in} \forall k,l=1,\dots, M,
\end{align}
where $||.||_{sp}$ denotes the spectral norm and $||.||_{F}$ denotes the Forbenius norm of a matrix.

\begin{figure*}[!t]
\normalsize
\setcounter{equation}{28}
\begin{align}
\label{real}
\Re(\boldsymbol{\alpha}^{H} \textbf{C}^{k,l} \boldsymbol{\alpha})&=[\Re(\boldsymbol{\alpha})^{T} \hspace{.1 in} \Im(\boldsymbol{\alpha})^{T}]\begin{bmatrix}
\ \Re(\textbf{C}^{k,l}) & &-\Im(\textbf{C}^{k,l}) \\ 
\ \Im(\textbf{C}^{k,l}) & &\Re(\textbf{C}^{k,l})    
\end{bmatrix}\begin{bmatrix}
\ \Re(\boldsymbol{\alpha}) \\
\ \Im(\boldsymbol{\alpha})    \\
\end{bmatrix} =[\Re(\boldsymbol{\alpha})^{T} \hspace{.1 in} \Im(\boldsymbol{\alpha})^{T}](\textbf{C}_{\Re}^{k,l})\begin{bmatrix}
\ \Re(\boldsymbol{\alpha}) \\
\ \Im(\boldsymbol{\alpha})    \\
\end{bmatrix}.
\end{align}
\begin{align}
\label{imaginary}
\Im(\boldsymbol{\alpha}^{H} \textbf{C}^{k,l} \boldsymbol{\alpha})&=[\Re(\boldsymbol{\alpha})^{T} \hspace{.1 in} \Im(\boldsymbol{\alpha})^{T}]\begin{bmatrix}
\ \Im(\textbf{C}^{k,l}) & &\Re(\textbf{C}^{k,l}) \\ 
\ -\Re(\textbf{C}^{k,l}) & &\Im(\textbf{C}^{k,l})    
\end{bmatrix}\begin{bmatrix}
\ \Re(\boldsymbol{\alpha}) \\
\ \Im(\boldsymbol{\alpha})    \\
\end{bmatrix} =[\Re(\boldsymbol{\alpha})^{T} \hspace{.1 in} \Im(\boldsymbol{\alpha})^{T}](\textbf{C}_{\Im}^{k,l})\begin{bmatrix}
\ \Re(\boldsymbol{\alpha}) \\
\ \Im(\boldsymbol{\alpha})    \\
\end{bmatrix}.
\end{align}
\setcounter{equation}{34}
\begin{align}
\label{cov_mat}
 \boldsymbol{\Theta}=\begin{bmatrix}
\ \frac{1}{4N}[Tr(\textbf{C}_{\Re}^{1,1}(\textbf{C}_{\Re}^{1,1}))+Tr(\textbf{C}_{\Re}^{1,1}(\textbf{C}_{\Re}^{1,1})^{T})]   & &\dots &  &\frac{1}{4N}[Tr(\textbf{C}_{\Re}^{1,1}(\textbf{C}_{\Im}^{M,M}))+Tr(\textbf{C}_{\Re}^{1,1}(\textbf{C}_{\Im}^{M,M})^{T})] \\
\ \frac{1}{4N}[Tr(\textbf{C}_{\Re}^{1,2}(\textbf{C}_{\Re}^{1,1}))+Tr(\textbf{C}_{\Re}^{1,2}(\textbf{C}_{\Re}^{1,1})^{T})]  &   &\dots  & &\frac{1}{4N}[Tr(\textbf{C}_{\Re}^{1,2}(\textbf{C}_{\Im}^{M,M}))+Tr(\textbf{C}_{\Re}^{1,2}(\textbf{C}_{\Im}^{M,M})^{T})] \\
\ \vdots & &\ddots & &\vdots \\
\ \frac{1}{4N}[Tr(\textbf{C}_{\Re}^{M,M}(\textbf{C}_{\Re}^{1,1}))+Tr(\textbf{C}_{\Re}^{M,M}(\textbf{C}_{\Re}^{1,1})^{T})]    &  &\dots &  &\frac{1}{4N}[Tr(\textbf{C}_{\Re}^{M,M}(\textbf{C}_{\Im}^{M,M}))+Tr(\textbf{C}_{\Re}^{M,M}(\textbf{C}_{\Im}^{M,M})^{T})]\\
\ \frac{1}{4N}[Tr(\textbf{C}_{\Im}^{1,1}(\textbf{C}_{\Re}^{1,1}))+Tr(\textbf{C}_{\Im}^{1,1}(\textbf{C}_{\Re}^{1,1})^{T})]  &  &\dots & &\frac{1}{4N}[Tr(\textbf{C}_{\Im}^{1,1}(\textbf{C}_{\Im}^{M,M}))+Tr(\textbf{C}_{\Im}^{1,1}(\textbf{C}_{\Im}^{M,M})^{T})] \\
\ \frac{1}{4N}[Tr(\textbf{C}_{\Im}^{1,2}(\textbf{C}_{\Re}^{1,1}))+Tr(\textbf{C}_{\Im}^{1,2}(\textbf{C}_{\Re}^{1,1})^{T})]  &  &\dots & &\frac{1}{4N}[Tr(\textbf{C}_{\Im}^{1,2}(\textbf{C}_{\Im}^{M,M}))+Tr(\textbf{C}_{\Im}^{1,2}(\textbf{C}_{\Im}^{M,M})^{T})] \\
\ \vdots & &\ddots & &\vdots \\
\ \frac{1}{4N}[Tr(\textbf{C}_{\Im}^{M,M}(\textbf{C}_{\Re}^{1,1}))+Tr(\textbf{C}_{\Im}^{M,M}(\textbf{C}_{\Re}^{1,1})^{T})]  &  &\dots & &\frac{1}{4N}[Tr(\textbf{C}_{\Im}^{M,M}(\textbf{C}_{\Im}^{M,M}))+Tr(\textbf{C}_{\Im}^{M,M}(\textbf{C}_{\Im}^{M,M})^{T})] \\
\end{bmatrix}.  
\end{align}
\hrulefill
\vspace*{4pt}
\end{figure*}

Therefore \textbf{H}$\textbf{H}^{H}$ is a $M \times M$ matrix of the following general form,
\setcounter{equation}{26}
\begin{eqnarray}
\label{forapp}
 \frac{1}{N}
\begin{bmatrix}
\ \boldsymbol{\alpha}^{H} \textbf{C}^{1,1} \boldsymbol{\alpha}   &\boldsymbol{\alpha}^{H} \textbf{C}^{1,2} \boldsymbol{\alpha} &\dots &\boldsymbol{\alpha}^{H}\textbf{C}^{1,M} \boldsymbol{\alpha}    \\
\ \boldsymbol{\alpha}^{H} \textbf{C}^{2,1} \boldsymbol{\alpha}    &\boldsymbol{\alpha}^{H}\textbf{C}^{2,2} \boldsymbol{\alpha} &\dots &\boldsymbol{\alpha}^{H}\textbf{C}^{2,M} \boldsymbol{\alpha}   \\
\ \vdots &\vdots &\ddots &\vdots \\
\ \boldsymbol{\alpha}^{H}\textbf{C}^{M,1} \boldsymbol{\alpha}    &\boldsymbol{\alpha}^{H}\textbf{C}^{M,2} \boldsymbol{\alpha} &\dots &\boldsymbol{\alpha}^{H}\textbf{C}^{M, M} \boldsymbol{\alpha}
\end{bmatrix},
\end{eqnarray}
where $\textbf{C}^{k,l} = [\bar{\textbf{b}}_{k}^{H} \bar{\textbf{b}}_{l}]\circ (\textbf{A}^{H} \textbf{A})^{T}$, $\forall k,l$ satisfy assumption \textbf{A-2} and are Hermitian, positive semi-definite matrices. This technical assumption, made to use the Central Limit Theorem developed for a quadratic form in normal random variables, serves as a means of ensuring that the matrix \textbf{A} is not rank deficient. This can be seen using Schur's classical inequality  $||[\bar{\textbf{b}}_{k}^{H} \bar{\textbf{b}}_{l}]\circ (\textbf{A}^{H} \textbf{A})^{T}||_{sp} \leq ||[\bar{\textbf{b}}_{k}^{H} \bar{\textbf{b}}_{l}] ||_{sp}  ||(\textbf{A}^{H} \textbf{A})^{T}||_{sp}$ coupled with Lemma 1 in \cite{gesbert_lemma}. The physical interpretation of this assumption is that the angles of departure that appear in the exponential terms of the array responses in \textbf{A} must be sufficiently separable to ensure that $\frac{||\textbf{A}||_{sp}}{||\textbf{A}||_{F}} \xrightarrow{} 0$. 

We investigate the asymptotic behaviour of the joint distribution of the $M^{2}$ entries of \textbf{H}$\textbf{H}^{H}$ under the assumptions \textbf{A-1} and \textbf{A-2}. Denoting the $(k,l)^{th}$ quadratic term by $T^{k,l}= \frac{1}{N} \boldsymbol{\alpha}^{H} \textbf{C}^{k,l}  \boldsymbol{\alpha}$, every quadratic term can be decomposed into its real and imaginary parts increasing the total number of terms that constitute $\textbf{H}\textbf{H}^{H}$ to $2M^{2}$. These parts are also quadratic forms in $\boldsymbol{\alpha}$ as shown below,
\begin{equation}
\begin{aligned}
T^{k,l}=\frac{1}{N}\Re(\boldsymbol{\alpha}^{H} \textbf{C}^{k,l} \boldsymbol{\alpha})+j\frac{1}{N}\hspace{.04 in} \Im(\boldsymbol{\alpha}^{H}  \textbf{C}^{k,l} \boldsymbol{\alpha}), \hspace{.08 in} \forall k,l=1, \dots , M,
\end{aligned}
\end{equation}
where $\Re(\boldsymbol{\alpha}^{H} \textbf{C}^{k,l} \boldsymbol{\alpha})$ and $\Im(\boldsymbol{\alpha}^{H} \textbf{C}^{k,l} \boldsymbol{\alpha})$ are given by (\ref{real}) and (\ref{imaginary}) respectively.

Equivalently \textbf{H}$\textbf{H}^{H}$ can be written as a sum of two matrices containing the real and imaginary parts of the quadratic terms.
\setcounter{equation}{30}
\begin{align}
& \textbf{H}\textbf{H}^{H}=\frac{1}{N}\begin{bmatrix}
\ \Re(\boldsymbol{\alpha}^{H} \textbf{C}^{1,1}\boldsymbol{\alpha})  & &\dots & &\Re(\boldsymbol{\alpha}^{H}\textbf{C}^{1,M}\boldsymbol{\alpha})    \\
\ \Re(\boldsymbol{\alpha}^{H} \textbf{C}^{2,1}\boldsymbol{\alpha}) &   &\dots & &\Re(\boldsymbol{\alpha}^{H}\textbf{C}^{2,M}\boldsymbol{\alpha})    \\
\ \vdots & &\ddots & &\vdots \\
\ \Re(\boldsymbol{\alpha}^{H}\textbf{C}^{M,1}\boldsymbol{\alpha})  & &\dots & &\Re(\boldsymbol{\alpha}^{H}\textbf{C}^{M,M}\boldsymbol{\alpha})
\end{bmatrix}  \nonumber \\
&+ \frac{j}{N} \begin{bmatrix}
\ \Im(\boldsymbol{\alpha}^{H} \textbf{C}^{1,1}\boldsymbol{\alpha})  &  &\dots & &\Im(\boldsymbol{\alpha}^{H}\textbf{C}^{1,M}\boldsymbol{\alpha})    \\
\ \Im(\boldsymbol{\alpha}^{H}\textbf{C}^{2,1}\boldsymbol{\alpha})  &  &\dots  & &\Im(\boldsymbol{\alpha}^{H}\textbf{C}^{2,M}\boldsymbol{\alpha})    \\
\ \vdots & &\ddots & &\vdots \\
\ \Im(\boldsymbol{\alpha}^{H}\textbf{C}^{M,1}\boldsymbol{\alpha}) & &\dots & &\Im(\boldsymbol{\alpha}^{H}\textbf{C}^{M,M}\boldsymbol{\alpha})
\end{bmatrix}. 
\end{align}

 Note that $\mathbb{E}[\Re(\alpha)^{2}]=\mathbb{E}[\Im(\alpha)^{2}]=\frac{1}{2}$. With this decomposition at hand, we now state in the following theorem the asymptotic behavior of the joint distribution of the entries of $\Re(\textbf{H}\textbf{H}^{H})$ and $\Im(\textbf{H}\textbf{H}^{H})$ stacked in vector \textbf{x} given by,
\begin{align}
\textbf{x}&=[\text{vec}(\Re(\textbf{H}\textbf{H}^{H}))^{T} \hspace{.04 in} \text{vec}(\Im(\textbf{H}\textbf{H}^{H}))^{T}]^{T},
\end{align}
where the operator vec(.) maps an $M \times M$ matrix to an $M^{2} \times 1$ vector by stacking the rows of the matrix. 

\begin{figure*}[!t]
\normalsize
\setcounter{equation}{37}
\begin{align}
\label{meanMI}
&\textbf{m}=[\Re(\zeta_{11})+\frac{1}{2N}Tr(\textbf{C}_{\Re}^{1,1}), \hspace{.1 in} \Re(\zeta_{12})+\frac{1}{2N}Tr(\textbf{C}_{\Re}^{1,2}), \dots, \Re(\zeta_{MM})+\frac{1}{2N}Tr(\textbf{C}_{\Re}^{M,M}), \nonumber  \\
& \Im(\zeta_{11})+\frac{1}{2N}Tr(\textbf{C}_{\Im}^{1,1}), \hspace{.1 in} \Im(\zeta_{12})+\frac{1}{2N}Tr(\textbf{C}_{\Im}^{1,2}), \dots, \Im(\zeta_{MM})+\frac{1}{2N}Tr(\textbf{C}_{\Im}^{M,M})]^{T}
\end{align}
\hrulefill
\vspace*{4pt}
\end{figure*}

\textit{Theorem 3:}
Let $\textbf{H}$ be a $N_{MS} \times N_{BS}$ MIMO channel matrix from (\ref{entrop}), then under the setting of assumptions \textbf{A-1} and \textbf{A-2}, \textbf{x} behaves as multivariate Gaussian such that the MGF of \textbf{x} has the following convergence behavior,
\setcounter{equation}{32}
\begin{align}
\label{Theorem3}
\mathbb{E}\left[\exp \left(\frac{\sqrt{N}\textbf{s}^{T}(\textbf{x}-\textbf{m})}{\sqrt{\textbf{s}^{T} \boldsymbol{\Theta} \textbf{s}}} \right) \right]-\exp\left(\frac{1}{2} \right) \xrightarrow{} 0,
\end{align}
where,
\begin{align}
\label{mean}
\textbf{m}&=\frac{1}{2N}[Tr(\textbf{C}_{\Re}^{1,1})  \dots  Tr(\textbf{C}_{\Re}^{M,M}), \hspace{.05 in} Tr(\textbf{C}_{\Im}^{1,1})  \dots Tr(\textbf{C}_{\Im}^{M,M})]^{T},
\end{align}
and $\boldsymbol{\Theta}$ is given by (\ref{cov_mat}).

The proof of Theorem 3 is postponed to Appendix A. Note that this theorem can be used in applications involving the statistical distribution of $\textbf{H}\textbf{H}^{H}$ other than the MI analysis.

\subsection{Distribution of Mutual Information}
After studying the asymptotic behaviour of $\Re(\textbf{H}\textbf{H}^{H})$, $\Im(\textbf{H}\textbf{H}^{H})$ for the proposed entropy maximizing channel, the CDF of the MI is worked out. To this end, note that (\ref{MI}) can be equivalently expressed as,
\setcounter{equation}{35}
\begin{align}
\label{MI_dec}
I(\sigma^{2})&=\log \text{det} (\textbf{R}+\sigma^2 \textbf{I}_{M}+ \textbf{H}\textbf{H}^{H})-\log \text{det}(\textbf{R}+\sigma^2 \textbf{I}_{M}).
\end{align}

Decomposing $((\textbf{R}+\sigma^2 \textbf{I}_{M})+ \textbf{H}\textbf{H}^{H})$ into real and imaginary parts as,
\begin{align}
\label{YZ}
& \Re((\textbf{R}+\sigma^2 \textbf{I}_{M})+ \textbf{H}\textbf{H}^{H})+ j \hspace{.05 in} \Im((\textbf{R}+\sigma^2 \textbf{I}_{M})+ \textbf{H}\textbf{H}^{H}), \nonumber \\
&= \textbf{Y}+j \hspace{.03 in} \textbf{Z},
\end{align}
and denoting the entries of $\textbf{R}+\sigma^{2}\textbf{I}_{M}$ by $\zeta_{ij}$, we can extend the result of Theorem 3 to the entries of $\Re((\textbf{R}+\sigma^2 \textbf{I}_{M})+ \textbf{H}\textbf{H}^{H})$ and $\Im((\textbf{R}+\sigma^2 \textbf{I}_{M})+ \textbf{H}\textbf{H}^{H})$, which will also show the convergence behavior in (\ref{Theorem3}) under assumptions \textbf{A-1} and \textbf{A-2} with mean vector $\textbf{m}$ given by (\ref{meanMI}) and the same covariance. 

The mean matrix \textbf{M} for $((\textbf{R}+\sigma^2 \textbf{I}_{M})+ \textbf{H}\textbf{H}^{H})$ is expressed as,

\setcounter{equation}{38}
\begin{align}
\label{M}
\textbf{M}&=\textbf{M}_{1}+j \hspace{.03 in} \textbf{M}_{2},
\end{align}
where,
\begin{align}
\label{M1}
& \textbf{M}_{1}=
\begin{bmatrix}
\ \Re(\zeta_{11})+\frac{1}{2N}Tr (C_{\Re}^{1,1})   &\dots  &\Re(\zeta_{1M})+\frac{1}{2N}Tr (C_{\Re}^{1,M})     \\
\ \Re(\zeta_{21})+\frac{1}{2N}Tr (C_{\Re}^{2,1})     &\dots  &\Re(\zeta_{2M})+\frac{1}{2N}Tr (C_{\Re}^{2,M})    \\
\ \vdots  &\ddots  &\vdots \\
\ \Re(\zeta_{M1})+\frac{1}{2N}Tr (C_{\Re}^{M,1})   &\dots  &\Re(\zeta_{MM}) +\frac{1}{2N}(C_{\Re}^{M,M}) 
\end{bmatrix} ,
\\
& \textbf{M}_{2}=
 \begin{bmatrix}
\ \Im(\zeta_{11})+\frac{1}{2N}Tr (C_{\Im}^{1,1})   &\dots  &\Im(\zeta_{1M})+\frac{1}{2N}Tr (C_{\Im}^{1,M})     \\
\ \Im(\zeta_{21})+\frac{1}{2N}Tr (C_{\Im}^{2,1})     &\dots  &\Im(\zeta_{2M})+\frac{1}{2N}Tr (C_{\Im}^{2,M})    \\
\ \vdots  &\ddots   &\vdots \\
\ \Im(\zeta_{M1})+\frac{1}{2N}Tr (C_{\Im}^{M,1})   &\dots  & \Im(\zeta_{MM}) +\frac{1}{2N}(C_{\Im}^{M,M})
\end{bmatrix} .
\end{align}

Before presenting the Gaussian approximation to the statistical distribution of the MI in the asymptotic limit, we define a matrix $\widetilde{\textbf{M}}$ and a $2M^{2} \times 1$ vector $\underline{\textbf{f}}(\widetilde{\textbf{M}})$ as,
\begin{align}
\widetilde{\textbf{M}}=\begin{bmatrix}
\ \textbf{M}_{1} &  &-\textbf{M}_{2}    \\
\ \textbf{M}_{2} &  &\textbf{M}_{1}    \end{bmatrix},
\end{align}
\begin{align}
& \underline{\textbf{f}}(\widetilde{\textbf{M}})|_{l+(k-1)M}=[\text{det}(\textbf{D}_{1})+\dots+\text{det}(\textbf{D}_{2M})]|_{l+(k-1)M}, \nonumber \\
& \hspace{.1 in} k=1, \dots, 2M, l=1, \dots, M,
\end{align}
where $\textbf{D}_{i}$ is identical to $\widetilde{\textbf{M}}$ except that the entries in the $i^{th}$ row are replaced by their derivatives with respect to $(k,l)^{th}$ entry of $\begin{bmatrix}
\ \textbf{M}_{1}    \\
\ \textbf{M}_{2}   \end{bmatrix}$. Every entry of the $2M^{2} \times 1$ vector, $ \underline{\textbf{f}}(\widetilde{\textbf{M}})$ will involve the sum of only two non-zero determinants as every $(k,l)^{th}$ entry occurs in only two rows. Additionally we make the following assumption,
\\
\textbf{Assumption A-3.} In the large $N_{BS}, N$ regime presented in \textbf{A-1}, 
\begin{align}
\lim \inf \underline{\textbf{f}}(\widetilde{\textbf{M}})^{T} \hspace{.03 in} \boldsymbol{\Theta} \hspace{.05 in} \underline{\textbf{f}}(\widetilde{\textbf{M}}) > 0,
\end{align}
where $\boldsymbol{\Theta}$ is the covariance matrix from (\ref{cov_mat}). This assumption allows us to take into account the higher order terms in the Taylor series expansion needed to compute the MI distribution as shown in Appendix B.  The physical interpretation is that it allows us to see the fluctuations in the distribution of the MI, since ignoring the higher order terms will just give us a first order deterministic result. 

With all these definitions at hand, we now present in the following theorem the Gaussian approximation to the statistical distribution of the MI in the asymptotic limit.

\textit{Theorem 4:} Let $\textbf{H}$ be a $N_{MS} \times N_{BS}$ MIMO channel matrix from (\ref{entrop}), such that the assumptions \textbf{A-1}, \textbf{A-2} and \textbf{A-3} hold, then the statistical distribution of $I(\sigma^{2})$ can be approximated as,
\begin{equation}
\label{Theorem4}
\begin{aligned}
\mathbb{P}[\sqrt{N}I(\sigma^2) \leq x ] - \frac{1}{2}\left(1+erf\left(\frac{x-\sqrt{N}\mu}{\sqrt{2\sigma_{a}^{2}}}\right)\right) \xrightarrow{} 0,
\end{aligned}
\end{equation}
where $erf$ is the error function, $\mu= 0.5 \log \det \widetilde{\textbf{M}} - \log \det (\textbf{R}+\sigma^2 \textbf{I}_{M})$ and $\sigma_{a}^{2}$ is given by,
\begin{equation}
\begin{aligned}\left(\frac{.5}{ \det \widetilde{\textbf{M}}}\right)^{2} \times \underline{\textbf{f}}(\widetilde{\textbf{M}})^{T} \hspace{.03 in} \boldsymbol{\Theta}  \hspace{.05 in} \underline{\textbf{f}}(\widetilde{\textbf{M}}),
\end{aligned}
\end{equation}

The proof of Theorem 4 is postponed to Appendix B. The approximation becomes more accurate as $N_{BS}$ and $N$ grow large but yields a good fit for even moderate values of $N_{BS}$ and $N$. The Gaussian approximation to the MI is an effective tool for the performance evaluation of future 3D MIMO channels in massive MIMO systems that will scale up the current MIMO systems by possible order of magnitudes. However accommodating more antennas at the MS introduces several constraints for practical implementations which would not affect the presented results since the approximation is valid for any number of Rx antennas

\section{Numerical Results}

 \begin{figure}[!t]
\centering
\includegraphics[width=2 in]{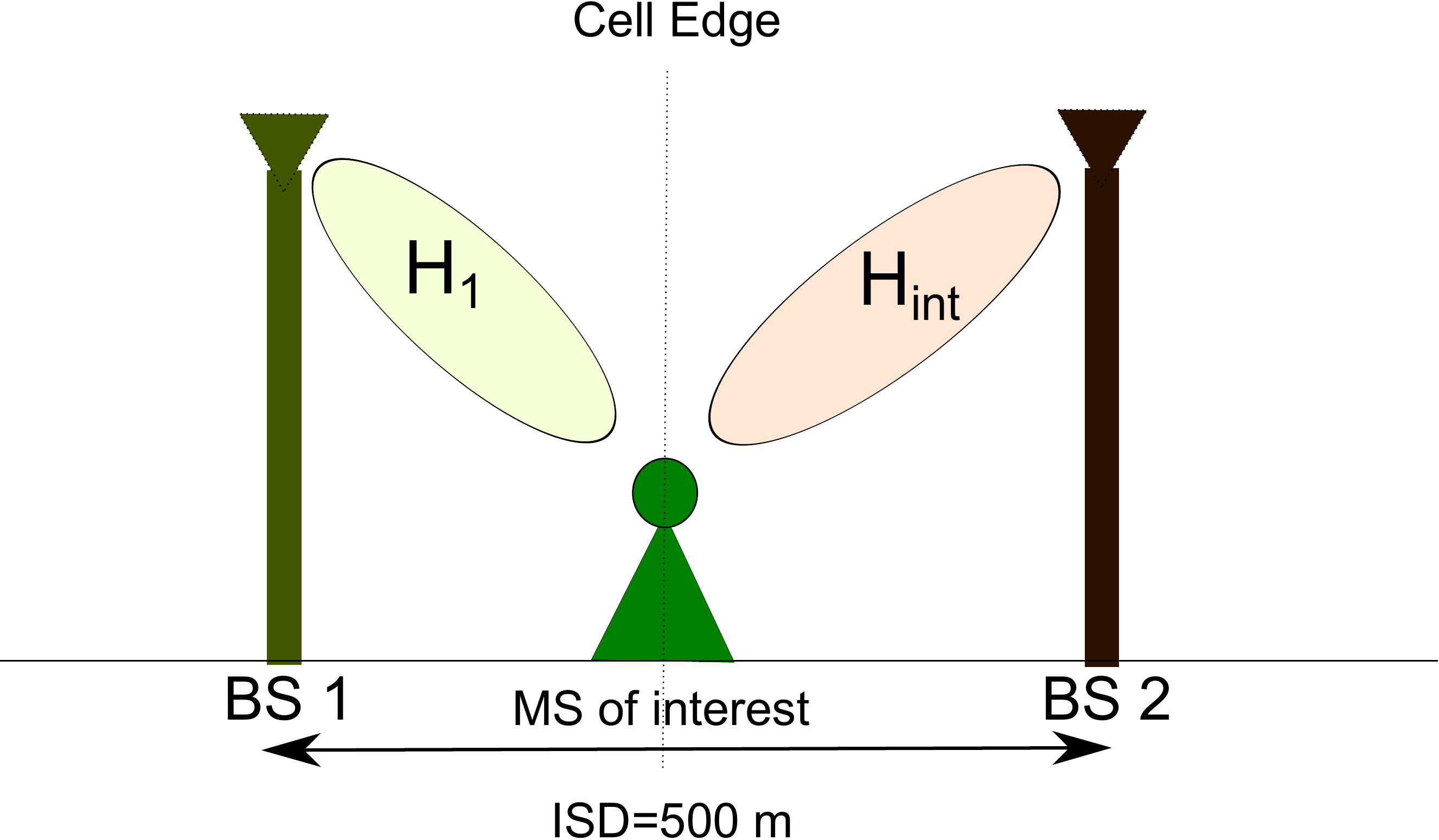}
\caption{Example scenario.}
 \label{scenario}
\end{figure}

\begin{figure}[!t]
\centering
\includegraphics[width=2.5 in]{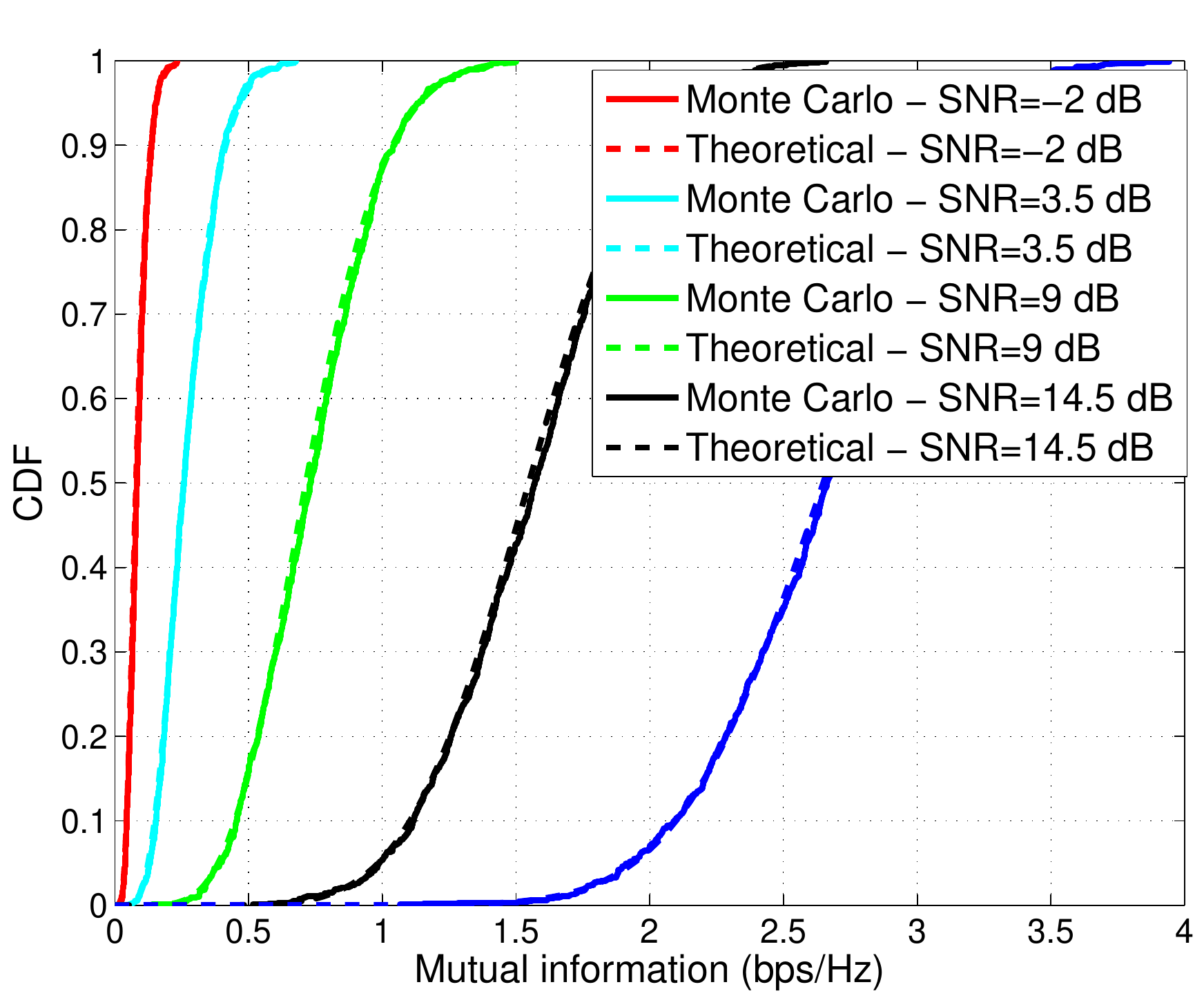}
\caption{Comparison of Monte Carlo simulated CDF and theoretical CDF derived in (\ref{Theorem2}) for the general SINR regime.}
\label{exact}
\end{figure}

We corroborate the results derived in this work with simulations. A simple but realistic scenario shown in Fig. \ref{scenario} is studied to help gain insight into the accuracy of the derived CDFs and the impact of the careful selection of antenna downtilt angles on the system MI in the presence of interfering BSs. This multi-cell scenario consists of a single MS and two BSs separated by an inter-site distance of 500m. The BS height is 25m and MS height is 1.5m. The MS is associated to BS 1 and can be located inside a cell of outer radius 250m and inner radius 35m. We will study the worst case performance when the MS is at the cell edge. The MI given in (\ref{MI}) requires the computation of the interference matrix.  The complex received baseband signal at the MS in the multi-cell scenario will be given as,
\begin{align}
\textbf{y}=\textbf{H}\textbf{x}+\sum_{i=1}^{N_{int}} \textbf{H}_{i} \textbf{x}_{i} + \textbf{n},
\end{align}
where $N_{int}$ is the number of interfering BSs ($N_{int}$=1 for the scenario in Fig. \ref{scenario} and \textbf{n} is the additive white Gaussian noise with variance $\sigma^{2}$. The channel between the serving BS and MS is denoted by \textbf{H} from (\ref{entrop}) and the channels $\textbf{H}_{i}$ between $N_{int}$ interfering BSs and MS follow a similar matrix representation.

The resultant interference matrix is then given by,
\begin{equation}
\begin{aligned}
\textbf{R}=\sum_{i=1}^{N_{int}} \mathbb{E}[\textbf{H}_{i}\textbf{H}_{i}^{H}],
\end{aligned}
\end{equation}
where,
\begin{equation}
\begin{aligned}
\label{interf}
[\textbf{H}_{i}\textbf{H}_{i}^{H}]=\frac{1}{N}\textbf{B}_{i}\text{diag}(\alpha)\textbf{A}_{i}^{H} \textbf{A}_{i} \text{diag}(\alpha^{*}) \textbf{B}_{i}^{H}, \hspace{.1 in} i=1,\dots, N_{int}.
\end{aligned}
\end{equation}

Since this paper largely focuses on the guidelines provided in the mobile communication standards used globally, so it is important that we use the angular distributions and antenna patterns specified in the standards for our analysis.  In the standards, elevation AoD and AoA are drawn from Laplacian elevation density spectrum given by,
\begin{equation}
\begin{aligned}
f_{\theta}(\theta) \propto \exp\left(-\frac{\sqrt{2}|\theta-\theta_{0}|}{\sigma}\right),
\end{aligned}
\end{equation}
where $\theta_{0}$ is the mean AoD/AoA and $\sigma$ is the angular spread in the elevation \cite{PES1}.  The characteristics of the azimuth angles are well captured by Wrapped Gaussian (WG) density spectrum \cite{ITU}, \cite{Kalliola02angularpower} which can be approximated quite accurately by the Von Mises (VM) distribution \cite{correlation2}, \cite{vonmises1} given by,
\begin{eqnarray}
f_{\phi}(\phi)=\frac{\exp(\kappa \cos(x-\mu))}{2\pi I_{0}(\kappa)},
\end{eqnarray}
where $I_{n}(\kappa)$ is the modified Bessel function of order $n$, $\mu$ is the mean AoD/AoA and $\frac{1}{\kappa}$ is a measure of azimuth dispersion. Since the channel assumes single-bounce scattering, so the angles are generated once and are assumed to be constant over the channel realizations.

\begin{figure}[!t]
\centering
\includegraphics[width=2.5 in]{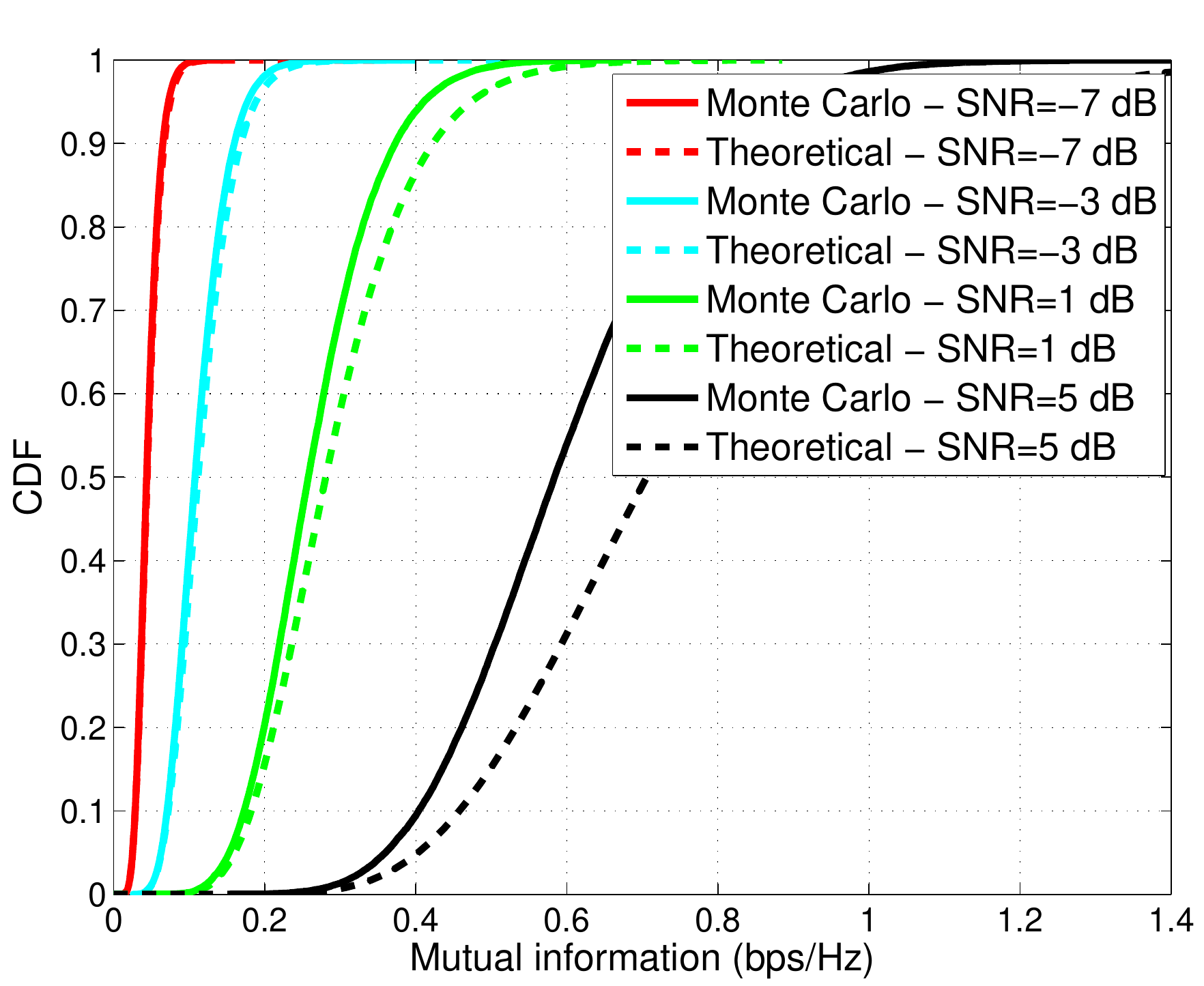}
\caption{Comparison of Monte Carlo simulated CDF and theoretical CDF derived in (\ref{Theorem1}) for the low SINR regime.}
\label{lowsnr}
\end{figure}
\begin{figure}
\centering
\includegraphics[width=2.5 in]{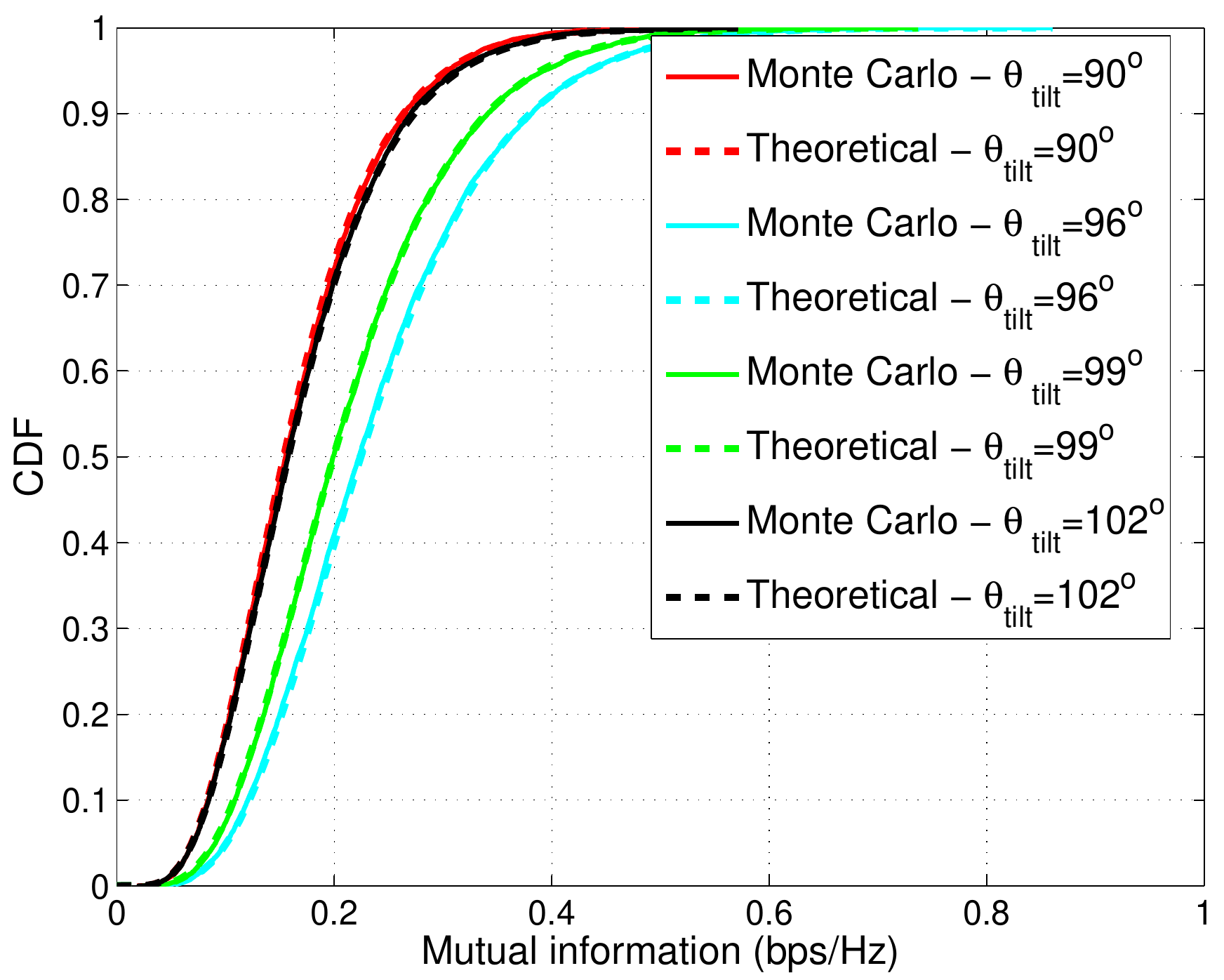}
\caption{Comparison in a multi-cell environment for different values of $\theta_{tilt}$.}
\label{multi}
\end{figure}

The radio and channel parameters are set as $\theta_{3dB}=15^{o}$, $\phi_{3dB}=70^{o}$, $\sigma_{BS}=7^{o}$, $\sigma_{MS}=10^{o}$, $\kappa_{BS}, \kappa_{MS}=5$ and $\mu=0$. Moreover $\theta_{0}$ is set equal to the elevation line of sight (LoS) angle ($\theta_{LoS}$) between the BS and MS. Denoting $\bigtriangleup h$ as the height difference between the MS and the BS, and defining $x$ and $y$ as the relative distance between the MS and the BS in the $x$ and $y$ coordinates respectively, the elevation LoS angle with respect to the horizontal plane at the BS can be written as $\theta_{LoS}=\frac{\pi}{2}-\tan^{-1}\frac{\bigtriangleup h}{\sqrt{x^{2}+y^{2}}}$. Two thousand Monte Carlo realizations of the 3D channel in (\ref{channel}) are generated to obtain the simulated MI for comparison.  In the simulations, $SNR$ is used to denote $\frac{1}{\sigma^{2}}$ and is given in $\rm{dB}$.

The closed-form expression of the CDF of MI provided in Theorem 1 (\ref{Theorem2}) for the general SINR regime is validated through simulation in Fig. \ref{exact} for $N_{BS}=20$ and $N_{MS}=1$. The number of propagation paths $N$ is assumed to be $40$ and $\theta_{tilt}$ is set equal to the elevation LoS angle of the BS with the MS. For this result, BS 2 is assumed to be operating in a different frequency band as BS 1 so the MS does not face any interference and $(\textbf{R}+\sigma^2 \textbf{I}_{N_{MS}})^{-1}$ in (\ref{MI}) reduces to $\frac{1}{\sigma^2} \textbf{I}_{N_{MS}}$. The closed-form expression for the CDF provides a perfect fit to the MI obtained using Monte Carlo realizations of the channel in (\ref{channel}). In the next numerical result, we validate the expression in (\ref{Theorem1}) for the low SINR regime. The result is plotted for $N_{MS}=2$ in Fig. \ref{lowsnr}. The approximation is seen to coincide quite well with the simulated results for the proposed 3D channel model at low values of $SNR$.  However for $SNR$ level of 5 $\rm{dB}$ or above, the fit is not very good.  The reason can be attributed to the approximation in (19) failing to hold. This can be seen by realizing that higher values of $SNR$, correspond to lower values of $\sigma^{2}$ that results in $(\textbf{R}+\sigma^{2} \textbf{I}_{N_{BS}})^{-1}$ to have a non-negligible norm. Hence, although the closed-form expression obtained in (\ref{Theorem1}) works for any arbitrary number of receive antennas as opposed to (\ref{Theorem2}) but this expression is accurate only for low values of $SNR$. 

In the next simulation, we investigate the impact of the downtilt angle of the transmitting cell on the performance of the users at the cell edge in a multi-cell scenario, when both BSs are transmitting in the same frequency band. The interference matrix \textbf{R} is computed from (\ref{interf}). Note from Fig. \ref{scenario} that the MS is in the direct boresight of the interfering cell and has $\theta_{LoS}$ of $95.37^{o}$ with both the serving and interfering BS. It can be clearly seen from Fig. \ref{multi} that the performance of the user at the cell edge is very sensitive to the value of the downtilt angle. The simulation result  illustrates that the MI of the system is maximized when the serving BS also sets the elevation boresight angles of its antennas equal to the elevation LoS angle with the MS, i.e. $\theta_{tilt} \approx 96^{o}$. We plot the CDF for the exact expression of MI for a single Rx antenna provided in (\ref{Theorem2}) for $SNR= 5 \rm{dB}$. The perfect fit of the theoretical result to the simulated result for the CDF of the MI is again evident. 

\begin{figure}[!t]
\centering
\includegraphics[width=2.5 in]{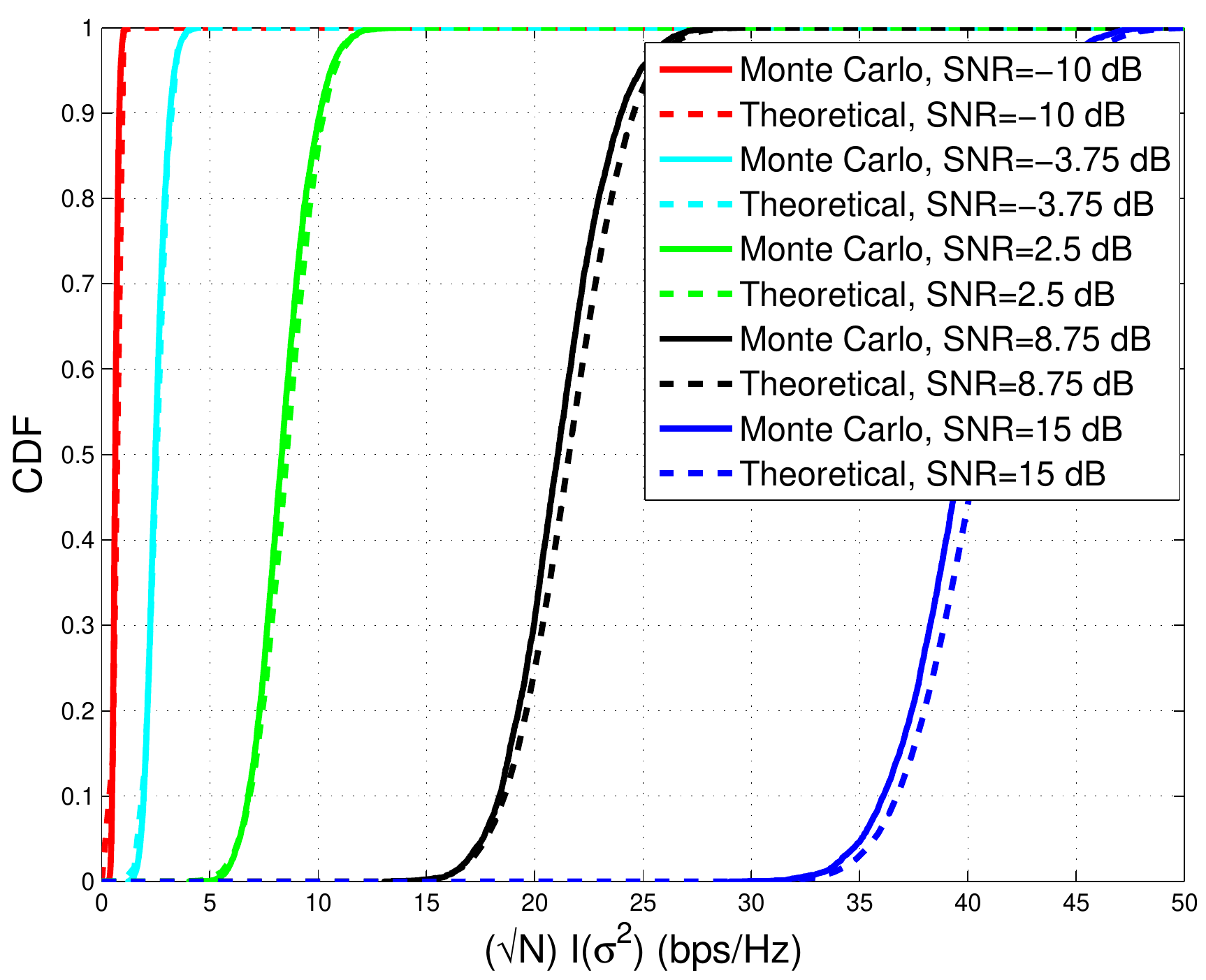}
\caption{Comparison of Monte Carlo simulated CDF and asymptotic theoretical CDF in (\ref{Theorem4}).}
\label{asym}
\end{figure}

\begin{figure}[!t]
\centering
\includegraphics[width=2.5 in]{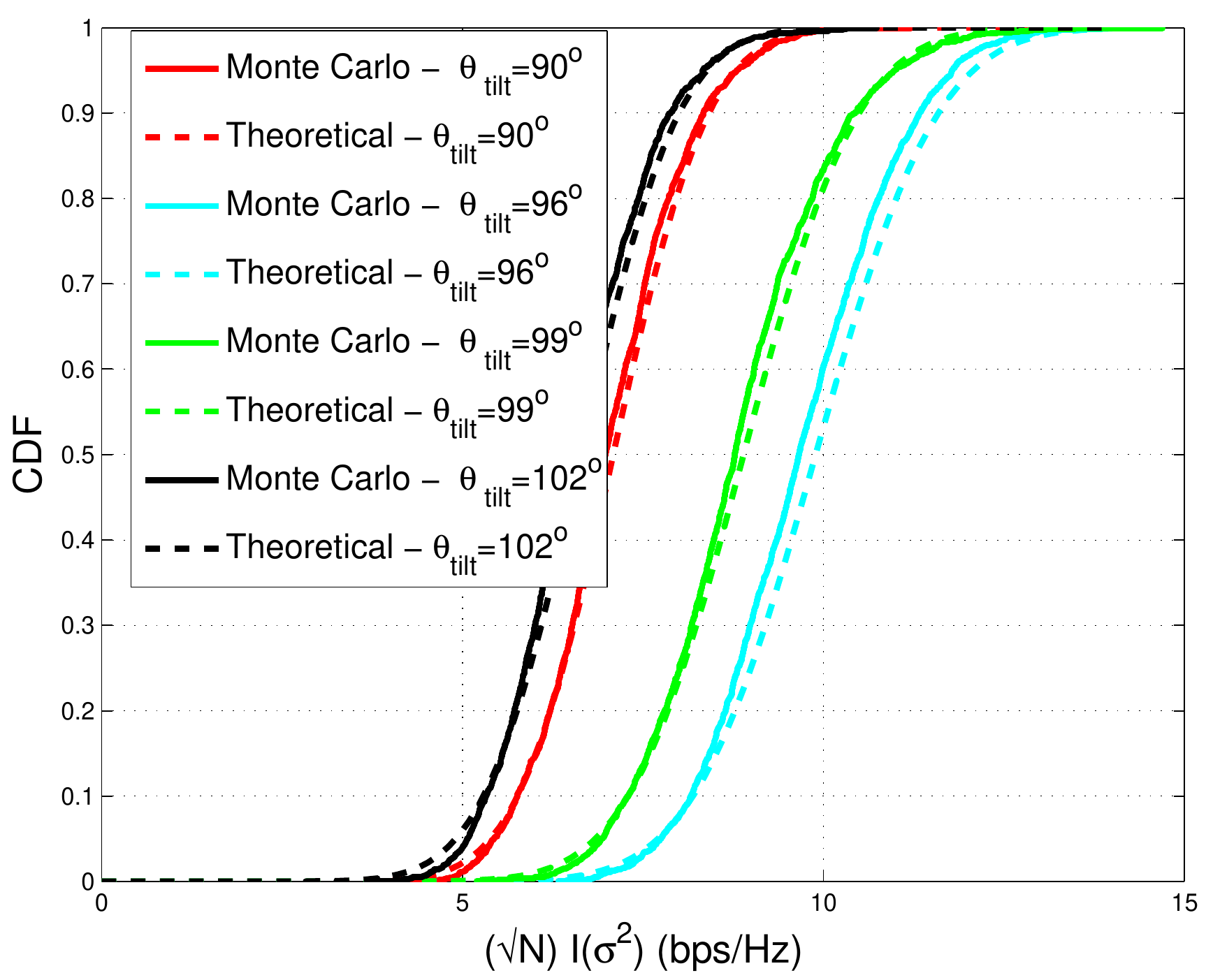}
\caption{Comparison of Monte Carlo simulated CDF and asymptotic theoretical CDF in a mutli-cell environment.}
\label{asym_multi}
\end{figure}

We now validate the result obtained for the CDF of MI in the asymptotic limit, as $N_{BS}$ and $N$ grow large, for the cases with finite values of $N_{BS}$ and $N$. The first result deals with scenario when the BS 2 is operating in a different frequency band as BS 1. Fig. \ref{asym} shows that there is quite a good fit between the asymptotic theoretical result and the simulated result for finite sized simulated system with $N_{BS}=60$, a reasonable number for large MIMO systems and $N_{MS}=4$. The number of paths is taken to be of the same order as $N_{BS}$. Even at the highest simulated value of $SNR$, the asymptotic mean and asymptotic variance show only a 1.2 and 3.2 percent relative error respectively. This close match illustrates the usefulness of the asymptotic approach in characterizing the MI for 3D maximum entropy channels for any number of Rx antennas. Hence, as far as performance analysis based on the achievable MI is considered, the asymptotic statistical distribution obtained can be used for the evaluation of future 3D beamforming strategies. Finally the multi-cell scenario is considered in Fig. \ref{asym_multi}, where we plot the CDF of the MI achieved by the proposed entropy maximizing channel in the asymptotic limit at different values of antenna boresight angles of the serving BS. We again deal with the worst case scenario of the MS being in the direct boresight of the interfering cell. Comparison with the simulated CDF of MI obtained at $SNR=5 \rm{dB}$ validates the accuracy of the derived theoretical asymptotic distribution and illustrates the impact 3D elevation beamforming can have on the system performance through the meticulous selection of downtilt angles. For the case on hand, the MI of the system is again maximized when the serving BS also sets its antenna boresight angles equal to the elevation LoS angle with the MS.

\begin{figure*}[!t]
\normalsize
\setcounter{equation}{52}
\begin{align}
\label{var}
Var[\Re(T^{k,l})]&=  \frac{1}{N^{2}}\mathbb{E}[|\Re(\alpha)|^{2}]^{2}[Tr(\textbf{C}^{k,l}_{\Re}\textbf{C}^{k,l}_{\Re})+Tr(\textbf{C}^{k,l}_{\Re}(\textbf{C}^{k,l}_{\Re})^{T})], \\
Var[\Im(T^{k,l})]&=  \frac{1}{N^{2}}\mathbb{E}[|\Im(\alpha)|^{2}]^{2}[Tr(\textbf{C}^{k,l}_{\Im}\textbf{C}^{k,l}_{\Im})+Tr(\textbf{C}^{k,l}_{\Im}(\textbf{C}^{k,l}_{\Im})^{T})]. \nonumber
\end{align}
\begin{align}
\label{cov}
Cov[\Re / \Im(T^{k,l})\Re / \Im(T^{k',l'})]&= \frac{1}{N^{2}} \mathbb{E}[|\Re/\Im(\alpha)|^{2}]^{2}[Tr(\textbf{C}^{k,l}_{\Re/\Im}\textbf{C}^{k',l'}_{\Re/\Im})+Tr(\textbf{C}^{k,l}_{\Re/\Im}(\textbf{C}^{k',l'}_{\Re/\Im})^{T})].
\end{align}
\setcounter{equation}{55}
\begin{align}
\label{xx}
\textbf{x}&=\frac{1}{N}[\Re(\boldsymbol{\alpha}^{H}\textbf{C}^{1,1}\boldsymbol{\alpha}) \hspace{.06 in} \dots \hspace{.06 in} \Re(\boldsymbol{\alpha}^{H} \textbf{C}^{M,M} \boldsymbol{\alpha}), \hspace{.1 in} \Im(\boldsymbol{\alpha}^{H} \textbf{C}^{1,1} \boldsymbol{\alpha}) \hspace{.06 in} \dots \hspace{.06 in} \Im(\boldsymbol{\alpha}^{H} \textbf{C}^{M,M} \boldsymbol{\alpha})]^{T}, \nonumber \\
\textbf{m}&=\frac{1}{2N}[Tr(\textbf{C}_{\Re}^{1,1}) \hspace{.06 in} \dots \hspace{.06 in} Tr(\textbf{C}_{\Re}^{M,M}),   \hspace{.1 in}  Tr(\textbf{C}_{\Im}^{1,1}) \hspace{.06 in} \dots \hspace{.06 in} Tr(\textbf{C}_{\Im}^{M,M})]^{T}.
\end{align}
\begin{align}
\label{mgfk}
\mathbb{E}\left[\exp\left(\textbf{g}^{T}[\textbf{x}-\textbf{m}]\right)\right]&=\mathbb{E}\left[\exp\left( \frac{1}{N} \begin{bmatrix}
\ \Re(\boldsymbol{\alpha})^{T}  & &\Im(\boldsymbol{\alpha})^{T}  
\end{bmatrix} \boldsymbol{\Xi} \begin{bmatrix}
\ \Re(\boldsymbol{\alpha}) \\
\ \Im(\boldsymbol{\alpha})    \\
\end{bmatrix} -\textbf{g}^{T}\textbf{m}\right)\right].
\end{align}
\setcounter{equation}{59}
\begin{align}
\label{cov_app}
& \boldsymbol{\Theta}=\begin{bmatrix}
\ \frac{1}{4N}[Tr(\textbf{C}_{\Re}^{1,1}(\textbf{C}_{\Re}^{1,1}))+Tr(\textbf{C}_{\Re}^{1,1}(\textbf{C}_{\Re}^{1,1})^{T})]    &\dots   &\frac{1}{4N}[Tr(\textbf{C}_{\Re}^{1,1}(\textbf{C}_{\Im}^{M,M}))+Tr(\textbf{C}_{\Re}^{1,1}(\textbf{C}_{\Im}^{M,M})^{T})] \\
\ \frac{1}{4N}[Tr(\textbf{C}_{\Re}^{1,2}(\textbf{C}_{\Re}^{1,1}))+Tr(\textbf{C}_{\Re}^{1,2}(\textbf{C}_{\Re}^{1,1})^{T})]     &\dots   &\frac{1}{4N}[Tr(\textbf{C}_{\Re}^{1,2}(\textbf{C}_{\Im}^{M,M}))+Tr(\textbf{C}_{\Re}^{1,2}(\textbf{C}_{\Im}^{M,M})^{T})] \\
\ \vdots &\ddots  &\vdots \\
\ \frac{1}{4N}[Tr(\textbf{C}_{\Re}^{M,M}(\textbf{C}_{\Re}^{1,1}))+Tr(\textbf{C}_{\Re}^{M,M}(\textbf{C}_{\Re}^{1,1})^{T})]       &\dots   &\frac{1}{4N}[Tr(\textbf{C}_{\Re}^{M,M}(\textbf{C}_{\Im}^{M,M}))+Tr(\textbf{C}_{\Re}^{M,M}(\textbf{C}_{\Im}^{M,M})^{T})]\\
\ \frac{1}{4N}[Tr(\textbf{C}_{\Im}^{1,1}(\textbf{C}_{\Re}^{1,1}))+Tr(\textbf{C}_{\Im}^{1,1}(\textbf{C}_{\Re}^{1,1})^{T})]       &\dots  &\frac{1}{4N}[Tr(\textbf{C}_{\Im}^{1,1}(\textbf{C}_{\Im}^{M,M}))+Tr(\textbf{C}_{\Im}^{1,1}(\textbf{C}_{\Im}^{M,M})^{T})] \\
\ \frac{1}{4N}[Tr(\textbf{C}_{\Im}^{1,2}(\textbf{C}_{\Re}^{1,1}))+Tr(\textbf{C}_{\Im}^{1,2}(\textbf{C}_{\Re}^{1,1})^{T})]      &\dots  &\frac{1}{4N}[Tr(\textbf{C}_{\Im}^{1,2}(\textbf{C}_{\Im}^{M,M}))+Tr(\textbf{C}_{\Im}^{1,2}(\textbf{C}_{\Im}^{M,M})^{T})] \\
\ \vdots &\ddots &\vdots \\
\ \frac{1}{4N}[Tr(\textbf{C}_{\Im}^{M,M}(\textbf{C}_{\Re}^{1,1}))+Tr(\textbf{C}_{\Im}^{M,M}(\textbf{C}_{\Re}^{1,1})^{T})]     &\dots   &\frac{1}{4N}[Tr(\textbf{C}_{\Im}^{M,M}(\textbf{C}_{\Im}^{M,M}))+Tr(\textbf{C}_{\Im}^{M,M}(\textbf{C}_{\Im}^{M,M})^{T})] \\
\end{bmatrix}  
\end{align}
\hrulefill
\vspace*{4pt}
\end{figure*}

\section{Conclusion}
The prospect of enhancing system performance through elevation beamforming has stirred a growing interest among researchers in wireless communications. A large effort is currently being devoted to produce accurate 3D channel models that will be of fundamental practical use in analyzing these elevation beamforming techniques. In this work, we proposed a 3D channel model that is consistent with the state of available knowledge on channel parameters, and is inspired from system level stochastic channel models that have appeared in standards like 3GPP, ITU and WINNER. The principle of maximum entropy was used to determine the distribution of the MIMO channel matrix based on the prior information on angles of departure and arrival. The resulting 3D channel model is shown to have a systematic structure that can be exploited to effectively derive the closed-form expressions for the CDF of the MI. Specifically, we characterized the CDF in the general and low SINR regimes. An exact closed-form expression for the general SINR regime was derived for systems with a single receive antenna. An asymptotic analysis, in the number of paths and BS antennas, of the achievable MI is also provided and validated for finite-sized systems via numerical results. The derived analytical expressions were shown to closely coincide with the simulated results for the proposed 3D channel model. An important observation made was that the meticulous selection of the downtilt angles at the transmitting BS can have a significant impact on the performance gains, confirming the potential of elevation beamforming. We believe that the results presented will enable a fair evaluation of the 3D channels being outlined in the future generation of mobile communication standards.

\appendices
\section{Proof of Theorem 3}
The mean and variance of every quadratic term that constitutes the real and imaginary parts of \textbf{H}$\textbf{H}^{H}$, i.e. $\Re(T^{k,l})$ and $\Im(T^{k,l})$ (note that these terms are formed as a consequence of the decomposition of the original matrix in (\ref{forapp}) into real and imaginary parts and hence are real quadratic terms), can be computed as given in (\ref{mean}) and (\ref{var}) respectively \cite{quadratic}.
\setcounter{equation}{51}
\begin{align}
\label{mean}
\mathbb{E}[\Re(T^{k,l})]&=\frac{1}{N}\mathbb{E}[|\Re(\alpha)|^{2}]  Tr(\textbf{C}^{k,l}_{\Re}), \\
\mathbb{E}[\Im(T^{k,l})]&=\frac{1}{N}\mathbb{E}[|\Im(\alpha)|^{2}]  Tr(\textbf{C}^{k,l}_{\Im}). \nonumber
\end{align}

Similarly, the covariance between two entries of $\Re(\textbf{H}\textbf{H}^{H})$ and $\Im(\textbf{H}\textbf{H}^{H})$ is given by (\ref{cov}). The proof for covariance is straightforward and can be done by realizing that if $X=\boldsymbol{\alpha}^{H} \textbf{A} \boldsymbol{\alpha}$ and $Y=\boldsymbol{\alpha}^{H} \textbf{B} \boldsymbol{\alpha}$ where $\boldsymbol{\alpha}$ and \textbf{A} and \textbf{B} are real then,
\begin{align}
& \mathbb{E}[ XY ]=\mathbb{E} [\sum_{i,j,k,l=1}^{N} \alpha_{i} \alpha_{j} \alpha_{k} \alpha_{l} \textbf{A}_{ij} \textbf{B}_{kl}] \nonumber \\
&= \sum_{i=1}^{N} \mathbb{E}[|\alpha_{i}|^{4}] \textbf{A}_{ii} \textbf{B}_{ii} + \sum_{i=j,k=l, i \neq k}^{N} \mathbb{E}[|\alpha_{i}|^{2}]^{2} \textbf{A}_{ii} \textbf{B}_{kk} \nonumber \\ &+ \sum_{i=k,j=l, i \neq j}^{N} \mathbb{E}[|\alpha_{i}|^{2}]^{2} \textbf{A}_{ij} \textbf{B}_{ij} + \sum_{i=l,j=k, i \neq j}^{N} \mathbb{E}[|\alpha_{i}|^{2}]^{2} \textbf{A}_{ij} \textbf{B}_{ji} \nonumber
\end{align} 
and the cumulant $(\mu_{4}-3 \mu_{2}^{2})$ for Gaussian RV's is 0.

Given the expressions for mean and covariances of quadratic forms in Gaussian RVs, the Central Limit Theorem for the $(k,l)^{th}$ entry of $\Re(\textbf{H}\textbf{H}^{H}$) or $\Im(\textbf{H}\textbf{H}^{H}$) can be established from the result in \cite{convergencequad} under assumptions \textbf{A-1} and \textbf{A-2}. Noting that $\mathbb{E}[|\Re(\alpha)|^{2}]=\mathbb{E}[|\Im(\alpha)|^{2}]=\frac{1}{2}$, this convergence implies that the MGF of $\Re(T^{k,l})$ has the following behavior,
\setcounter{equation}{54}
\begin{align}
\label{MGF}
\mathbb{E}\left[\exp\left(\frac{s \sqrt{N}[\Re(T^{k,l})-\frac{1}{2N}Tr(C^{k,l}_{\Re})]}{\sqrt{\frac{[Tr(\textbf{C}^{k,l}_{\Re}\textbf{C}^{k,l}_{\Re})+Tr(\textbf{C}^{k,l}_{\Re}(\textbf{C}^{k,l}_{\Re})^{T})]}{4N}}}\right)\right]-\exp\left(\frac{s^{2}}{2}\right) \xrightarrow{} 0.
\end{align}
The MGF of $\Im(T^{k,l})$ has the same behavior with $\Re$ replaced with $\Im$.

We stack the entries of $\Re(\textbf{H}\textbf{H}^{H})$, $\Im(\textbf{H}\textbf{H}^{H})$ and their means in $\textbf{x}$ and $\textbf{m}$ respectively given by (\ref{xx}). From \cite{convergencequad}, each term in $\textbf{x}$ is asymptotically Normal. To characterize the joint distribution of the entries of $\Re(\textbf{H}\textbf{H}^{H})$, $\Im(\textbf{H}\textbf{H}^{H})$, the behaviour of their linear combination, i.e. $\textbf{g}^{T}[\textbf{x}-\textbf{m}]$, where $\textbf{g}$ is any arbitrary vector, is studied. The MGF of this linear combination is given by (\ref{mgfk}), where $\boldsymbol{\Xi}=[g_{1}\textbf{C}_{\Re}^{1,1}+...+g_{M^{2}}\textbf{C}_{\Re}^{M,M}+g_{M^{2}+1}\textbf{C}_{\Im}^{1,1}+...+g_{2M^{2}}\textbf{C}_{\Im}^{M,M}]$. Note that ${\begin{bmatrix}
\ \Re(\boldsymbol{\alpha})^{T}  &\Im(\boldsymbol{\alpha})^{T}  
\end{bmatrix} \boldsymbol{\Xi} \begin{bmatrix}
\ \Re(\boldsymbol{\alpha}) \\
\ \Im(\boldsymbol{\alpha})    \\
\end{bmatrix}}$ is also a quadratic form in Normal RVs ($\Re(\boldsymbol{\alpha})$, $\Im(\boldsymbol{\alpha})$). Extending the convergence result in (\ref{MGF}) for a quadratic form in Normal RVs with $s=1$, $\mathbb{E}[|\Re(\alpha)|^{2}]=\mathbb{E}[|\Im(\alpha)|^{2}]=\frac{1}{2}$, we have
\setcounter{equation}{57}
\begin{eqnarray}
\label{R1}
\mathbb{E}\left[\exp\left(\frac{\sqrt{N}\textbf{g}^{T}[\textbf{x}-\textbf{m}]}{\sqrt{\frac{1}{4N}[Tr(\boldsymbol{\Xi}\boldsymbol{\Xi})+Tr(\boldsymbol{\Xi}(\boldsymbol{\Xi})^{H})]}}\right)\right]- \exp\left(\frac{1}{2}\right) \xrightarrow{}0,
\end{eqnarray}
where,
\begin{eqnarray}
\label{R2}
\frac{1}{4N}[Tr(\boldsymbol{\Xi}\boldsymbol{\Xi})+Tr(\boldsymbol{\Xi}(\boldsymbol{\Xi})^{H})] &{}={}& \textbf{g}^{T}\boldsymbol{\Theta} \textbf{g},
\end{eqnarray}
and $\boldsymbol{\Theta}$ is the $2M^{2}\times 2M^{2}$ covariance matrix of the entries of vector $\sqrt{N}\textbf{x}$ given by (\ref{cov_app}). The entries of $\boldsymbol{\Theta}$ are obtained using (\ref{cov}). Therefore,
\setcounter{equation}{60}
\begin{equation}
\begin{aligned}
\mathbb{E}\left[\exp \left(\frac{\sqrt{N} (\textbf{g}^{T}\textbf{x}-\textbf{g}^{T}\textbf{m})}{\sqrt{\textbf{g}^{T}\boldsymbol{\Theta} \textbf{g}}}\right)\right]-\exp \left(\frac{1}{2}\right) \xrightarrow{}0.
\end{aligned}
\end{equation}

\begin{figure*}[!t]
\normalsize
\setcounter{equation}{61}
\begin{align}
\label{xn_app}
\textbf{x}_{n}&=[\Re(\zeta_{11})+\frac{1}{N}\Re(\boldsymbol{\alpha}^{H}\textbf{C}^{1,1}\boldsymbol{\alpha})  \dots  \Re(\zeta_{MM})+\frac{1}{N}\Re(\boldsymbol{\alpha}^{H} \textbf{C}^{M,M} \boldsymbol{\alpha}), \Im(\zeta_{11})+\frac{1}{N}\Im(\boldsymbol{\alpha}^{H} \textbf{C}^{1,1} \boldsymbol{\alpha})  \dots  \Im(\zeta_{MM})+\frac{1}{N}\Im(\boldsymbol{\alpha}^{H} \textbf{C}^{M,M} \boldsymbol{\alpha})]^{T}
\end{align}
\setcounter{equation}{67}
\begin{align}
\label{tran1}
\boldsymbol{\bigtriangledown} f_{1}\left(\begin{bmatrix}
\ \textbf{M}_{1}   &-\textbf{M}_{2}    \\
\ \textbf{M}_{2}   &\textbf{M}_{1}    \end{bmatrix}\right) \Big|_{l+(k-1)M}&=[\text{det}(\textbf{D}_{1})+\dots+\text{det}(\textbf{D}_{2M})]|_{l+(k-1)M},
\end{align}
\setcounter{equation}{69}
\begin{equation}
\label{var_fin}
\begin{aligned}\left(\frac{.5}{ \text{det}\left(\begin{bmatrix}
\ \textbf{M}_{1}   &-\textbf{M}_{2}    \\
\ \textbf{M}_{2}   &\textbf{M}_{1}    \end{bmatrix}\right)}\right)^{2} \times \boldsymbol{\bigtriangledown}^{T} f_{1}\left(\begin{bmatrix}
\ \textbf{M}_{1}   &-\textbf{M}_{2}    \\
\ \textbf{M}_{2}   &\textbf{M}_{1}    \end{bmatrix}\right)  \boldsymbol{\Theta} \boldsymbol{\bigtriangledown} f_{1}\left(\begin{bmatrix}
\ \textbf{M}_{1}   &-\textbf{M}_{2}    \\
\ \textbf{M}_{2}   &\textbf{M}_{1}    \end{bmatrix}\right),
\end{aligned}
\end{equation}
\hrulefill
\end{figure*}

Since $\textbf{g}^{T}\textbf{x}$ behaves as a Gaussian RV with mean $\textbf{g}^{T}\textbf{m}$ and  variance given by $\frac{1}{N}\textbf{g}^{T}\boldsymbol{\Theta} \textbf{g}$, so it can be deduced that $\textbf{x}$ behaves as multivariate Gaussian distribution with mean $\textbf{m}$ and covariance $\frac{1}{N}\boldsymbol{\Theta}$. Therefore under assumptions \textbf{A-1} and \textbf{A-2}, the joint distribution of the entries of $\sqrt{N}([\text{vec}(\Re(\textbf{H}\textbf{H}^{H}))^{T}$, $\text{vec}(\Im(\textbf{H}\textbf{H}^{H}))^{T}]^{T} -\textbf{m})$ behaves approximately as multivariate Gaussian with mean $0$ and covariance matrix given by $\boldsymbol{\Theta}$. This completes the proof of Theorem 3.

\section{Proof of Theorem 4}

To complete the characterization of MI, the $\log \det$ transformation needs to be applied on the entries of \textbf{Y} and \textbf{Z}, i.e. $\Re((\textbf{R}+\sigma^2 \textbf{I}_{M})+ \textbf{H}\textbf{H}^{H})$ and $\Im((\textbf{R}+\sigma^2 \textbf{I}_{M})+ \textbf{H}\textbf{H}^{H})$ respectively. These entries can be stacked in vector $\textbf{x}_{n}$ as shown in (\ref{xn_app}).

To obtain the statistical distribution of MI from the joint distribution of the entries of $\textbf{x}_{n}$, we succumb to the Taylor series expansion of a real-valued function $f(\textbf{x}_{n})$, 
\setcounter{equation}{62}
\begin{align}
f(\textbf{x}_{n})&=f(\textbf{m})+\boldsymbol{\bigtriangledown}^{T} f(\textbf{m})(\textbf{x}_{n}-\textbf{m})+R_{n},
\end{align}
where \textbf{m} is the mean vector in (\ref{meanMI}) and $R_{n}$ = $o(||\textbf{x}_{n}-\textbf{m}||^{2})$. We observe that $o(||\textbf{x}_{n}-\textbf{m}||^{2})$ is $o(1)$ because $\textbf{x}_{n} -\textbf{m}\xrightarrow{p} 0$. Any mapping of $\textbf{x}_{n}-\textbf{m}$ will therefore be $o(1)$ which implies that $R_{n}=o(1)$. 
\begin{align}
\label{TaylorMGF}
\mathbb{E}[\exp\left(f(\textbf{x}_{n})-f(\textbf{m})\right)]=\mathbb{E}\left[\exp\left(\boldsymbol{\bigtriangledown}^{T} f(\textbf{m})(\textbf{x}_{n}-\textbf{m})\right)\right] + o(1).
\end{align}

Note that $\mathbb{E}[\exp\left(\boldsymbol{\bigtriangledown}^{T} f(\textbf{m})(\textbf{x}_{n}-\textbf{m})\right)]$ is analogous to $\mathbb{E}\left[\exp\left(\textbf{g}^{T}[\textbf{x}_{n}-\textbf{m}]\right)\right]$ with $\textbf{g}=\boldsymbol{\bigtriangledown} f(\textbf{m})$. Now using the convergence result in (\ref{R1}) with (\ref{R2}) and invoking Slutsky's Theorem, given that \\ $\lim \inf \boldsymbol{\bigtriangledown}^{T} f(\textbf{m}) \boldsymbol{\Theta} \boldsymbol{\bigtriangledown} f(\textbf{m}) > 0$ results in,
\begin{align}
\label{int_app}
&\mathbb{E}\left[\exp\left(\frac{\sqrt{N}[f(\textbf{x}_{n})-f(\textbf{m})]}{\sqrt{ \boldsymbol{\bigtriangledown}^{T} f(\textbf{m}) \boldsymbol{\Theta} \boldsymbol{\bigtriangledown} f(\textbf{m})}}\right)\right]-\exp \left(\frac{1}{2}\right) \xrightarrow{}0.
\end{align}

This is analogous to applying Delta Theorem on the entries of $\textbf{Y}+j \textbf{Z}$. Therefore for $f(\textbf{Y}+j \textbf{Z})=\log \det (\textbf{Y}+j \textbf{Z})$, we can say that $\sqrt{N} (\log \det (\textbf{Y}+j\textbf{Z}) - \log \det (\textbf{M}_{1} + j \textbf{M}_{2}))$ behaves as a Gaussian RV under assumptions \textbf{A-1} and \textbf{A-2}, with mean $0$, and variance given by $(\boldsymbol{\bigtriangledown}^{T} f(\textbf{M})\boldsymbol{\Theta} \boldsymbol{\bigtriangledown}f(\textbf{M}))$.

Define a function $h(\textbf{x}_{n})=f_{2} \circ f_{1} (\textbf{x}_{n})$, where $f_{1}(\textbf{x}_{n})=(\det (\textbf{Y}+ j \textbf{Z}))^{2}$ and $f_{2}(x)=0.5 \log x$. For this function the gradient is, 
\begin{align}
\boldsymbol{\bigtriangledown} h|_{\textbf{x}_{n}=\textbf{m}}= f_{2}'(\det(\textbf{M}_{1}+j \textbf{M}_{2})^{2}) \times \boldsymbol{\bigtriangledown} f_{1}|_{\textbf{Y}=\textbf{M}_{1}, \textbf{Z}=\textbf{M}_{2}}.
\end{align}
We use the following to determine the expression for $\boldsymbol{\bigtriangledown} f_{1}$,
\begin{align}
(\text{det}(\textbf{Y}+j\textbf{Z}))^{2} &=\text{det}\left(\begin{bmatrix}
\ \textbf{Y}   &\textbf{-Z}    \\
\ \textbf{Z}   &\textbf{Y}    \end{bmatrix}\right).
\end{align}

Therefore the ${(l+(k-1)M)}^{th}, k=1, \dots 2M, l=1, \dots M$ entry of the vector $\boldsymbol{\bigtriangledown} f_{1}$ is given by (\ref{tran1}) [Ch 6, \cite{determinant}], where $\textbf{D}_{i}$ is identical to $\begin{bmatrix}
\ \textbf{M}_{1}   &-\textbf{M}_{2}    \\
\ \textbf{M}_{2}   &\textbf{M}_{1}    \end{bmatrix}$ except that the entries in the $i^{th}$ row are replaced by their derivatives with respect to $(k,l)^{th}$ entry of $\begin{bmatrix}
\ \textbf{M}_{1}    \\
\ \textbf{M}_{2}   \end{bmatrix}$, $k=1, \dots, 2M, l=1, \dots, M$. Every entry of $ \boldsymbol{\bigtriangledown} f_{1}\left(\begin{bmatrix}
\ \textbf{M}_{1}   &-\textbf{M}_{2}    \\
\ \textbf{M}_{2}   &\textbf{M}_{1}    \end{bmatrix} \right)$ will involve the sum of only two non-zero determinants as every $(k,l)^{th}$ entry occurs in only two rows and,
\setcounter{equation}{68}
\begin{equation}
\begin{aligned}
f_{2}'(\det(\textbf{M}_{1}+j\textbf{M}_{2})^{2})=\frac{.5}{ \det \left(\begin{bmatrix}
\ \textbf{M}_{1}   &-\textbf{M}_{2}    \\
\ \textbf{M}_{2}   &\textbf{M}_{1}    \end{bmatrix}\right)}.
\end{aligned}
\end{equation}

Therefore from (\ref{int_app}), in the asymptotic limit, the distribution of $\sqrt{N} \Big( \log \det ((\textbf{R}+\sigma^2 \textbf{I}_{M})$  $+ \textbf{H}\textbf{H}^{H}) $ \\ $ - \frac{1}{2}\log \text{det}\left(\begin{bmatrix}
\ \textbf{M}_{1}   &-\textbf{M}_{2}    \\
\ \textbf{M}_{2}   &\textbf{M}_{1}    \end{bmatrix}\right) \Big)$ can be approximated by a Gaussian distribution with mean $0$ and variance given by (\ref{var_fin}), under the assumption that $\lim \inf \boldsymbol{\bigtriangledown}^{T} f_{1}\left(\begin{bmatrix}
\ \textbf{M}_{1}   &-\textbf{M}_{2}    \\
\ \textbf{M}_{2}   &\textbf{M}_{1}    \end{bmatrix}\right)  \boldsymbol{\Theta}  \boldsymbol{\bigtriangledown} f_{1}\left(\begin{bmatrix}
\ \textbf{M}_{1}   &-\textbf{M}_{2}    \\
\ \textbf{M}_{2}   &\textbf{M}_{1}    \end{bmatrix}\right) > 0$. Using this result together with (\ref{MI_dec}), we can establish the statistical distribution of MI in the asymptotic limit for 3D MIMO channels. This completes the proof of Theorem 4.

\bibliographystyle{IEEEtran}
\bibliography{bib}

\begin{IEEEbiography}[{\includegraphics[width=1 in, height=1.25 in,clip,keepaspectratio ]{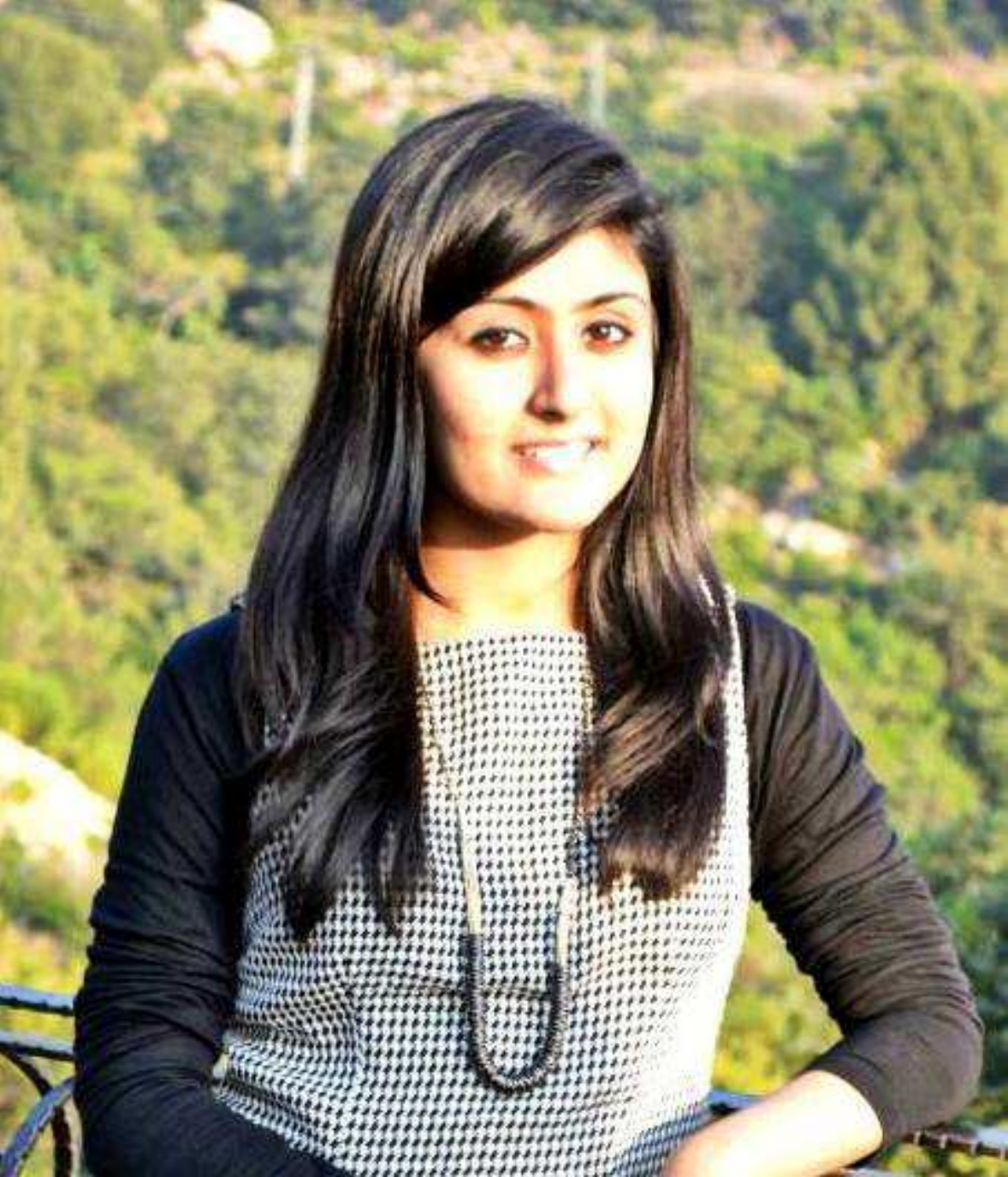}}]{Qurrat-Ul-Ain Nadeem}
Qurrat-Ul-Ain Nadeem was born in Lahore, Pakistan. She received her B.S. degree in electrical engineering from the Lahore University of Management Sciences (LUMS), Pakistan in 2013. She joined King Abdullah University of Science and Technology (KAUST), Thuwal, Makkah Province, Saudi Arabia in August 2013, where she is currently a M.S./Ph.D. student. Her research interests include channel modeling and performance analysis of wireless communications systems. 
\end{IEEEbiography}

\begin{IEEEbiographynophoto}{Abla Kammoun}
Abla Kammoun was born in Sfax, Tunisia. She received the engineering degree in signal and systems from the Tunisia Polytechnic School, La Marsa, and the Master's degree and the Ph.D. degree in digital communications from Telecom Paris Tech [then Ecole Nationale Sup{\'e}rieure des T{\'e}l{\'e}communications (ENST)]. From June 2010 to April 2012, she has been a Postdoctoral Researcher in the TSI Department, Telecom Paris Tech. Then she has been at Sup{\'e}lec at the Large Networks and Systems Group (LANEAS) until December 2013. Currently, she is  research scientist at KAUST university. Her research interests include performance analysis, random matrix theory, and semi-blind channel estimation.
\end{IEEEbiographynophoto}


\begin{IEEEbiographynophoto}{M{\'e}rouane Debbah}
M{\'e}rouane Debbah entered the Ecole Normale Sup{\'e}rieure de Cachan (France) in 1996 where he received his M.Sc and Ph.D. degrees respectively. He worked for Motorola Labs (Saclay, France) from 1999-2002 and the Vienna Research Center for Telecommunications (Vienna, Austria) until 2003. From 2003 to 2007, he joined the Mobile Communications department of the Institut Eurecom (Sophia Antipolis, France) as an Assistant Professor. Since 2007, he is a Full Professor at Sup{\'e}lec (Gif-sur-Yvette, France). From 2007 to 2014, he was director of the Alcatel-Lucent Chair on Flexible Radio. Since 2014, he is Vice-President of the Huawei France R\&D center and director of the Mathematical and Algorithmic Sciences Lab. His research interests lie in fundamental mathematics, algorithms, complex systems analysis and optimization, statistics, information \& communication sciences research. He is an Associate Editor in Chief of the journal Random Matrix: Theory and Applications and was an associate and senior area editor for IEEE Transactions on Signal Processing respectively in 2011-2013 and 2013-2014. M{\'e}rouane Debbah is a recipient of the ERC grant MORE (Advanced Mathematical Tools for Complex Network Engineering). He is a IEEE Fellow, a WWRF Fellow and a member of the academic senate of Paris-Saclay. He is the recipient of the Mario Boella award in 2005, the 2007 IEEE GLOBECOM best paper award, the Wi-Opt 2009 best paper award, the 2010 Newcom++ best paper award, the WUN CogCom Best Paper 2012 and 2013 Award, the 2014 WCNC best paper award as well as the Valuetools 2007, Valuetools 2008, CrownCom2009 , Valuetools 2012 and SAM 2014 best student paper awards. In 2011, he received the IEEE Glavieux Prize Award and in 2012, the Qualcomm Innovation Prize Award.
\end{IEEEbiographynophoto}

\begin{IEEEbiography}[{\includegraphics[width=1 in, height=1.25 in,clip,keepaspectratio ]{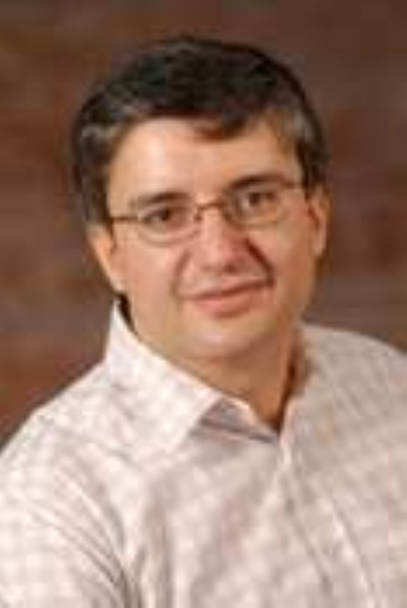}}]{Mohamed-Slim Alouini} (S'94, M'98, SM'03, F’09) Mohamed-Slim Alouini was born in Tunis, Tunisia. He received the Ph.D. degree in Electrical Engineering from the California Institute of Technology (Caltech), Pasadena, CA, USA, in 1998. He served as a faculty member in the University of Minnesota, Minneapolis, MN, USA, then in the Texas A\&M University at Qatar, Education City, Doha, Qatar before joining King Abdullah University of Science and Technology (KAUST), Thuwal, Makkah Province, Saudi Arabia as a Professor of Electrical Engineering in 2009. His current research interests include the modeling, design, and performance analysis of wireless communication systems.
\end{IEEEbiography}

\end{document}